\documentclass[%
 reprint,
 prab,
 amsmath,amssymb,
 aps,
floatfix,
]{revtex4-2}
\usepackage{graphicx}
\usepackage{dcolumn}
\usepackage{bm}
\usepackage{comment}
\usepackage{appendix}
\usepackage{subcaption}
\begin{document}

\title{Point-to-Point Coulomb Effects in High Brightness Photoelectron Beamlines for Ultrafast Electron Diffraction}
\author{ M. Gordon\textsuperscript{1}, S.B. van der Geer\textsuperscript{2}, J. Maxson\textsuperscript{3}, Y.-K. Kim\textsuperscript{1}, \\
\textsuperscript{1}University of Chicago Department of Physics, Chicago, IL, 60637; \\
\textsuperscript{2} Pulsar Physics, Burghstraat 47, 5614 BC Eindhoven, The Netherlands\\
\textsuperscript{3}Cornell University Department of Physics, Ithaca NY 14853}

\begin{abstract}

In an effort to increase spatial and temporal resolution of ultrafast electron diffraction and microscopy, ultra-high brightness photocathodes are actively sought to improve electron beam quality. Beam dynamics codes often approximate the Coulomb interaction with mean-field space charge, which is a good approximation in traditional beams. However, point-to-point Coulomb effects, such as disorder induced heating and the Boersch effect, can not be neglected in cold, dense beams produced by such photocathodes. In this paper, we introduce two new numerical methods to calculate the important effects of the photocathode image charge when using a point-to-point interaction model. Equipped with an accurate model of the image charge,  we calculate the effects of point-to-point interactions on two high brightness photoemission beamlines for ultrafast diffraction. The first beamline uses a 200 keV gun, whereas the second uses a 5 MeV gun, each operating in the single-shot diffraction regime with $10^5$ electrons/pulse. Assuming a zero photoemission temperature, it is shown that including stochastic Coulomb effects increases the final emittance of these beamlines by over a factor of 2 and decreases the peak transverse phase space density by over a factor of 3 as compared to mean-field simulations. We then introduce a method to compute the energy released by disorder induced heating using the pair correlation function. This disorder induced heating energy was found to scale very near the theoretical result for stationary ultracold plasmas, and it accounts for over half of the emittance growth above mean-field simulations.

\end{abstract}

\maketitle
\section{Introduction}

The development of high brightness photocathodes is a driving force in the improvement of electron accelerator technologies such as free electron lasers, energy recovery linacs, and ultrafast electron diffraction and microscopy (UED and UEM).  The brightness of the beams used in these applications is set at the electron source and can only degrade during further acceleration and transport. Consequently, the brightness of the electron source defines the ultimate limits of the capabilities of these devices  \cite{ADVCES,ERL_S,XFEL_S,UED_S}. The photocathode brightness is set by two parameters: the density of electrons emitted from the source, and their mean transverse energy (MTE), which acts as an effective beam temperature \cite{IvanThermalEnergy,IvanMTE}.  Increasing the electron density at the source is not always a viable option, as space charge forces can reduce brightness downstream dramatically. While some of this brightness can be restored via emittance compensation, some is lost to nonlinear distortions which are challenging to reverse \cite{EmitComp,EmitComp2}.   However, it has been shown in many modern applications that reducing the MTE of the photocathode can still lead to large gains in brightness \cite{CPierce}.  

Reducing the MTE of photocathodes is a very active area of research in which significant progress has been made in the last decade.  Typical photocathodes used in accelerator facilities today have a MTE of a few hundred meV \cite{SLAC_UED, LCLS, EXFEL} whereas near threshold emission at room temperature has demonstrated electron beams with an MTE of $\sim 25$ meV \cite{ThermalEmit}. Furthermore, cryocooled photocathodes near threshold have shown the capability to go down even further to an MTE of $\sim 5$ meV \cite{5meVMTE}. At these low temperatures, point-to-point interactions play an increasingly important role in the overall beam dynamics, as shown by the following argument.

The mean-field approximation commonly used in simulation codes is only valid when there are many particles in a Debye sphere. This Debye screening length is given by: 
\begin{equation}
    \lambda = \sqrt{\frac{\epsilon_0 k T}{n_0 e^2}  }
\end{equation}
where $\epsilon_0$ is the permittivity of free space, k is the Boltzmann constant, $T$ is the temperature of the beam, $n_0$ is the volume number density of the beam, and $e$ is the charge of the electron.  For $kT = 5$ meV and a density of $10^{17}\text{m}^{-1/3}$ (commonly achieved in photoinjectors today), the Debye screening length is approximately 1.7 $\mu$m.  However, the average interparticle spacing, $n_0^{-1/3}$ at the same density is 2.2 $\mu$m.  Thus very few electrons will be within one Debye screening length of any given electron \cite{Reiser}. This situation has been studied extensively for ultracold gas-based plasma and electron sources, which exhibit single meV electron temperatures in photoemission \cite{Scholten1,Scholten2,Luiten0,Luiten1,Luiten2,Luiten3,Luiten4,Luiten5}.

Brute force calculation of the pairwise Coulomb interaction scales with the square of the number of electrons, $\mathcal{O}(N^2)$, making it prohibitively time consuming to exactly simulate dynamics with large number of electrons.  Thus, to accurately capture the beam dynamics, approximation methods are used which  compute pairwise interactions of nearby particles, while approximating long range interactions using the mean-field approach.  These methods scale as $\mathcal{O}(N \log N)$ for traditional tree-based methods and $\mathcal{O}(N)$ for the fast multipole method, making them  feasible for simulation \cite{BHA, FMM}.  We will refer to these methods as point-to-point methods.

A critical challenge in employing a classical point-charge force model for a photoelectron source is the unphysical divergence of the image potential at the cathode surface. The underlying cause of the problem is that classical equations of motion are not valid at and very near the cathode surface.  In a classical simulation however, the size of the integration step typically scales inversely proportional to the gradient of the potential.  Thus, near a divergence, the integration step can limit to zero. This produces a scale-matching problem wherein very small step sizes  must be maintained throughout the particle emission process, which can lead to prohibitively long simulation times. However, as will be shown below, image charge effects significantly impact beam size and emittance evolution, and cannot be ignored.

This work aims to extend the work on Coulomb effects in ultracold plasma electron sources to photocathode guns, which can potentially support even higher beam density.  We provide a new method to compute the image force which is free of divergences and tuning parameters. Using this model of the overall beam dynamics, we turn to introduce new microscopic figures of merit to disentangle the global and local effects of point-to-point interactions.

To show the generality of the new methods, we examine beam dynamics in two very different UED beamlines based on archetypes used in practice today: a 200 keV dc gun with lower total initial beam density ($\sim 10^{17} \text{ m}^{-3}$), and a high gradient 5 MeV RF  photoinjector with higher initial beam density ($\sim 10^{18} \text{ m}^{-3}$).  UED is a good test case for examining point-to-point effects  as the number of particles needed per bunch is often relatively small ($10^5$-$10^7$) in comparison to synchrotron radiation applications, making it feasible to simulate every particle in the bunch with modest computing resources.  Along with this, the short bunch lengths, small spot sizes, and long coherence lengths needed to make atomic scale resolution diffraction patterns with femtosecond time resolution ultimately result in peak current densities comparable to those in free electron laser injectors  \cite{ADVCES,Sciaini_2011}.

Using our new method of calculating near-photocathode dynamics, we highlight one unique point-to-point phenomenon called disorder induced heating (DIH) which arises very near the photocathode. DIH was originally studied in the ultracold plasma community (see e.g.  \cite{ucpdih}), but it may have significant implications for cold photoemitted electron beams  \cite{Luiten2,Jared_DIH}. DIH is the thermalization of the initial potential energy stored in the random positions of near neighbor photoelectrons. Upon thermalization, the particle distribution develops a characteristic microscopic structure with a lack of near neighbors, and the beam simultaneously suffers emittance growth due to the increased temperature. For a stationary (non-accelerating, no expansion) electron bunch starting with zero temperature, the temperature rise due to DIH is given by:
\begin{equation}
k T_{\rm DIH} = \frac{C e^2}{4\pi \epsilon_0 a}
\label{eq:DIH}
\end{equation}
where $a=(3/4\pi n_0)^{1/3}$ is the Wigner-Seitz radius of the bunch, and $k$ is the Boltzmann constant. $C$ is a dimensionless constant which can be determined by tabulated plasma correlation energies to be roughly $C\approx 0.45$ \cite{TrappedNonneutralPlasma,Jared_DIH}. The timescale $\tau$ of the thermalization can be calculated to be a constant fraction of a plasma period \cite{Chandra}. In photoinjectors, this time is typically of the order of ten picoseconds or less, resulting in thermalization near the photocathode during initial acceleration. A correct image charge model is thus critical to understand DIH in photoinjectors. Furthermore, the beam density can change significantly near the photocathode due to space charge expansion and acceleration. The resulting balance has not been studied in detail in photocathode guns before. In this work we quantify DIH via an analysis of the resulting microscopic density distribution. We then estimate the rms emittance increase attributable to DIH, and find that it is the dominant source of emittance dilution in the two cases under study.

\section{Point-to-Point Simulation Methods}

\subsection{Coulomb interactions and Image Charge Model}
The simulations shown in this paper were performed with the space charge tracking code General Particle Tracer (GPT) \cite{ref:gpt_website} using three different algorithms. The first is GPT's mean-field space charge algorithm, a non-equidistant 3D multi-grid Poisson solver \cite{ref:gpt_MeanFieldAlgorithm}, which is used to calculate the mean-field interaction of the entire bunch, including image charge effects. The second is the Barnes-Hut algorithm internal to GPT \cite{BHA}, with a Barnes-Hut angle parameter of 0.3, used to model point-to-point Coulomb interactions.  However, due to divergent fields at the cathode, the Barnes-Hut algorithm does not allow for simple inclusion of image effects.  To include these effects, we developed a third technique, which we call the Plus-Minus-Plus (PMP) Method.

\begin{figure}[htbp]
\centering
\includegraphics[width=220pt]{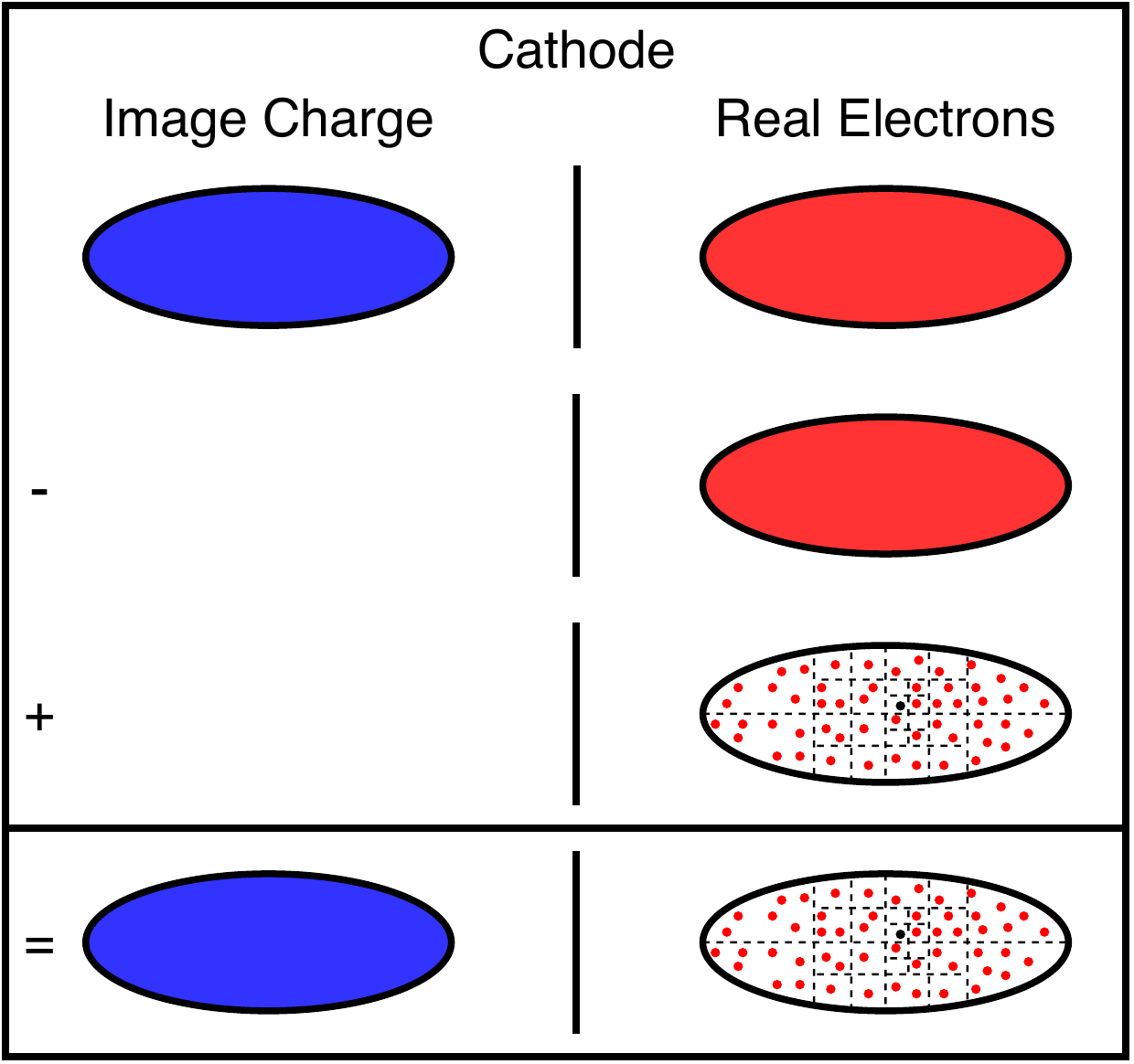}
\caption{Depiction of PMP 3-step space charge calculation.  Filled in ellipses represent mean-field calculation of electric fields, and ellipses filled with dots represent a Barnes-Hut calculation of electric fields.}  
\label{fig:PMP}
\end{figure}

The PMP method approximates the image charge as arising from a mean-field calculation and the total space charge force is calculated in a three step process as depicted in Fig. \ref{fig:PMP}. In PMP, GPT's mean-field space charge algorithm is used to calculate the mean-field interaction of the entire bunch, including image effects.
Subsequently, a second call is made to the mean-field solver, but this time without the cathode boundary condition. By subtracting this field from the initial full mean-field space charge calculation, the mean-field approximation of the image field alone is extracted.  The last step in the PMP process adds the stochastic interactions to the previously obtained mean-field image charges with the stochastic Barnes-Hut point-to-point method.

In an appendix we discuss the accuracy of the assumption of a mean-field image force. Specifically we compare the PMP method with another image charge method which includes point-to-point effects via a dynamical image charge potential which does not diverge \cite{DImC}. The latter requires additional computing time and tuning parameters, and in general we find good agreement with PMP. Thus PMP is our method of choice throughout. 

To achieve an accurate accounting of stochastic Coulomb effects, each macroparticle represents exactly 1 electron in all simulations, and all distributions are pseduo-random, rather than quasi-random.

\subsection{Using 90\% RMS Figures of Merit}

Root-mean-squared (rms) figures of merit for beam size/length, energy spread, and emittance are not well-defined in the presence of strong point-to-point Coulomb interactions. This is due to the presence of large angle scattering which generates long-tailed distributions for which rms values diverge \cite{JANSEN1990496}. To avoid sensitivity to outliers, but retain the sense of the traditional accelerator figures of merit, all quantities presented in this paper are calculated using 90\% rms values unless otherwise denoted, wherein a subset of the distribution containing 90\% of the particles are chosen such that the metric in question is minimized.

\section{Description of dc and NCRF gun UED beamlines}

Both beamline designs considered here originated from a multiobjective genetic algorithm (MOGA) optimization study, using the mean-field space charge model, to provide an emittance minimum at the sample plane with realistic constraints on the bunch length and spot size. These optimizations are described in Refs. \cite{CPierce,dc_Beamline_Optimization, RF_Beamline_Optimization}. Each case shown here is an individual from a MOGA Pareto optimal frontier. Individuals were selected containing a charge closest to $10^5$ electrons/pulse, and were then re-evalutated using the PMP method.    

\begin{figure}[htp]
\centering
\includegraphics[width=200pt]{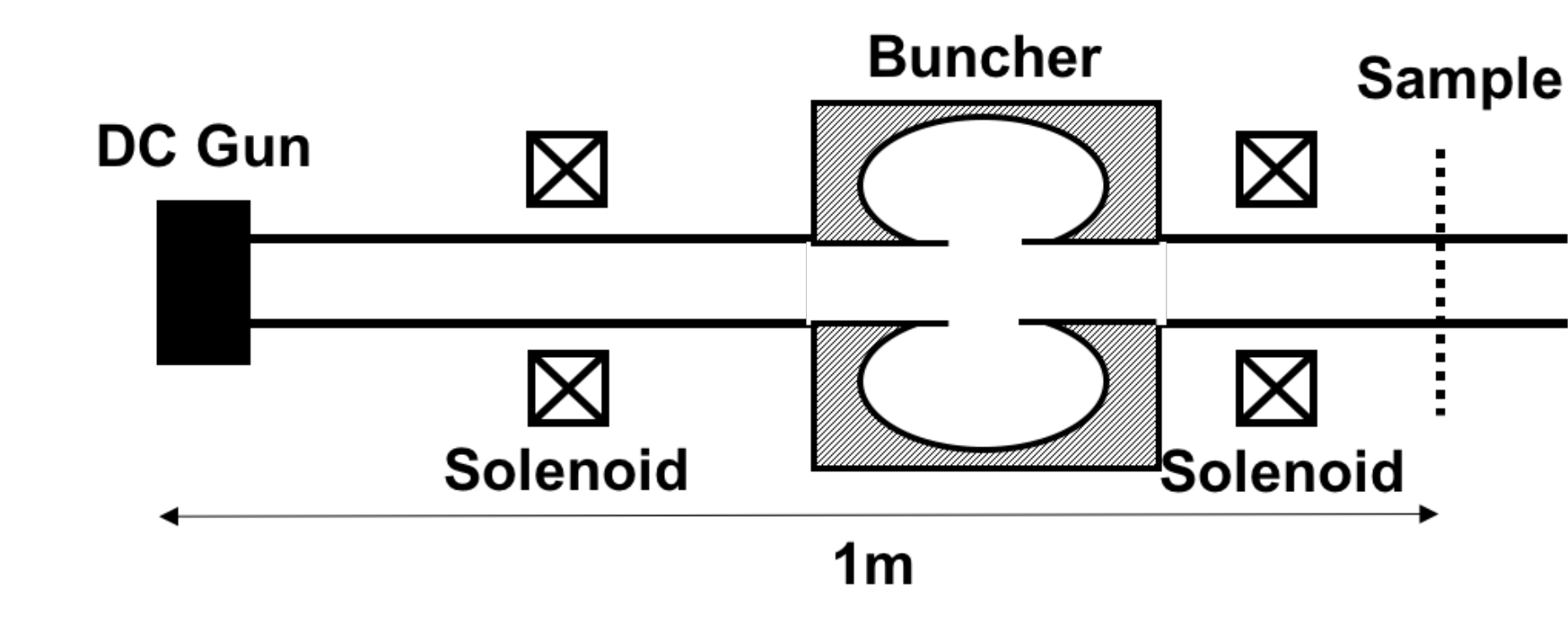}
\caption{Layout of the cryocooled dc gun UED beamline used in the following simulations.}
\label{fig:dcG} 
\end{figure}

The lower energy dc beamline, depicted schematically in Fig. \ref{fig:dcG}, consists of a cryocooled 200 kV dc gun \cite{CCdcGUN} with an extraction electric field of 11.25 MV/m, followed by a solenoid, a normal conducting 3.0 GHz buncher cavity of the Eindhoven design \cite{M_TM010}, and a second solenoid.  The beamline was optimized to have a minimal emittance at a sample location approximately 1 m from the cathode when emitting an electron beam with 0 meV MTE \cite{dc_Beamline_Optimization}.  The bunch charge is $-14$ fC.  The initial 100\% transverse rms size of the beam is 8.1 $\mu$m and the 100\% rms laser pulse length is 9.8 ps.

\begin{figure}[htbp]
\centering
\includegraphics[width=200pt]{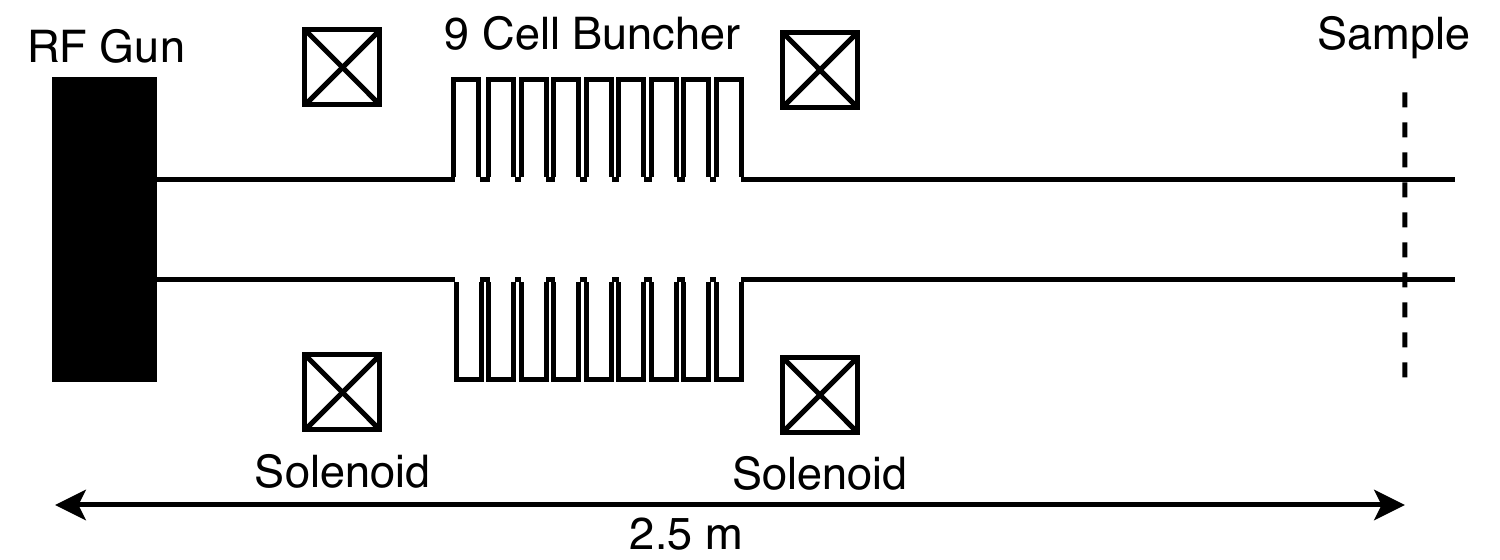}
\caption{Layout of the NCRF gun beamline used in the following simulations.}
\label{fig:RFG} 
\end{figure}

The higher energy NCRF beamline, depicted schematically in Fig. \ref{fig:RFG}, consists of a 1.6 cell 2.856 GHz NCRF gun of the BNL/SLAC/UCLA design \cite{RFGUN}, with a peak electric field of 100 MV/m, launch phase of 38.6 degrees from peak field, and final beam energy of 5 MeV, followed by a solenoid, a 9 cell buncher cavity, and a second solenoid.  The buncher is modeled using 9 copies of the first cell of the SLAC linac \cite{SlacCav}.  The beamline was optimized to have a minimal emittance at a sample location approximately 2.5 m from the cathode when emitting an electron beam with 0 meV MTE \cite{RF_Beamline_Optimization}.  The bunch charge used in the simulation is $-17$ fC. The initial 100\% transverse rms size of the beam is 2.5 $\mu$m and the 100\% rms laser pulse length is 3.2 ps.

\section{Macroscopic Beam Evolution}
\begin{figure}

\begin{subfigure}[htbp]{0.45\textwidth}
\centering
\includegraphics[width=\linewidth]{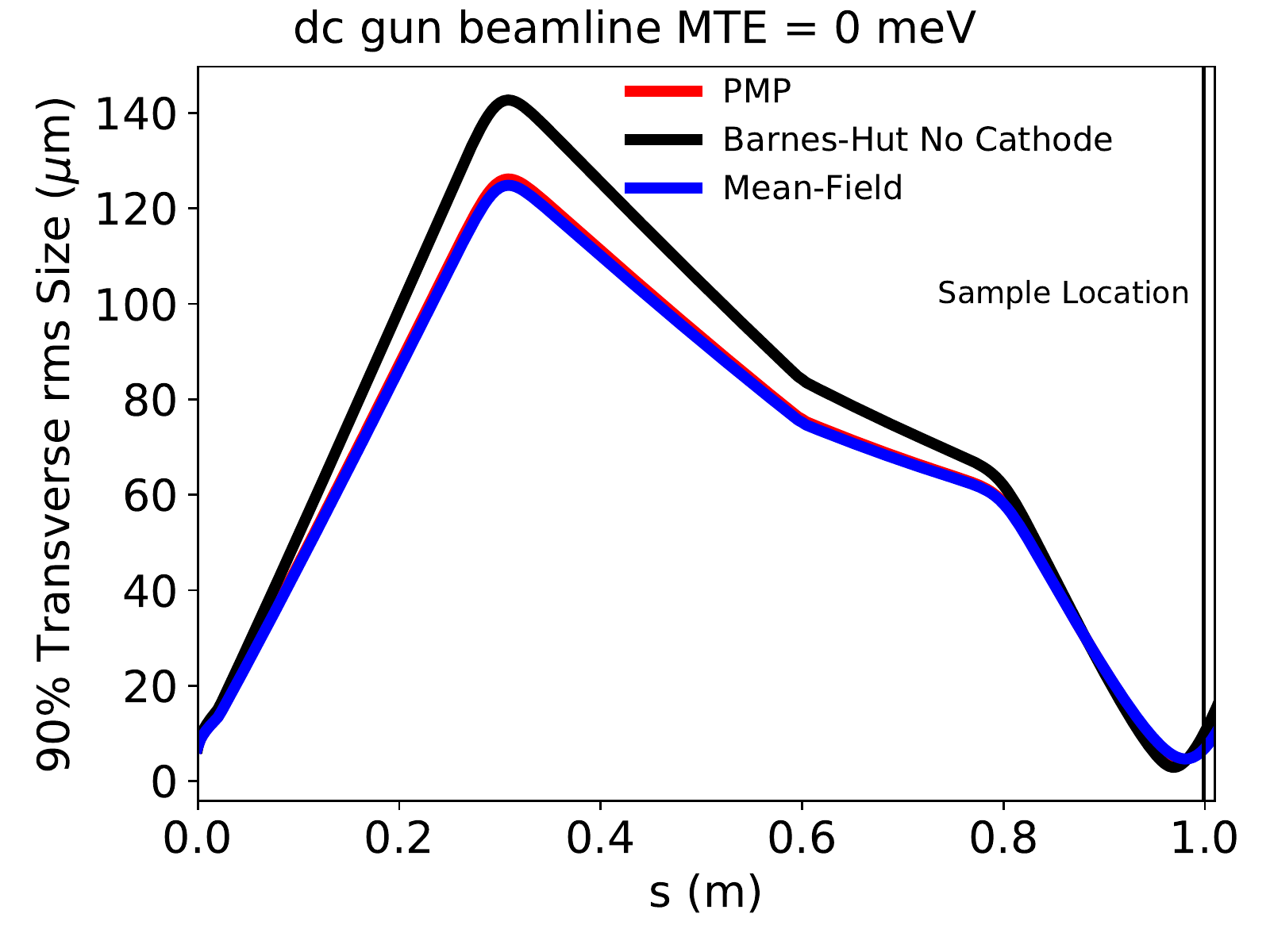}
\caption{}
\label{fig:NC_CdcSS}
\end{subfigure}

\begin{subfigure}[htp]{0.45\textwidth}
\centering
\includegraphics[width=\linewidth]{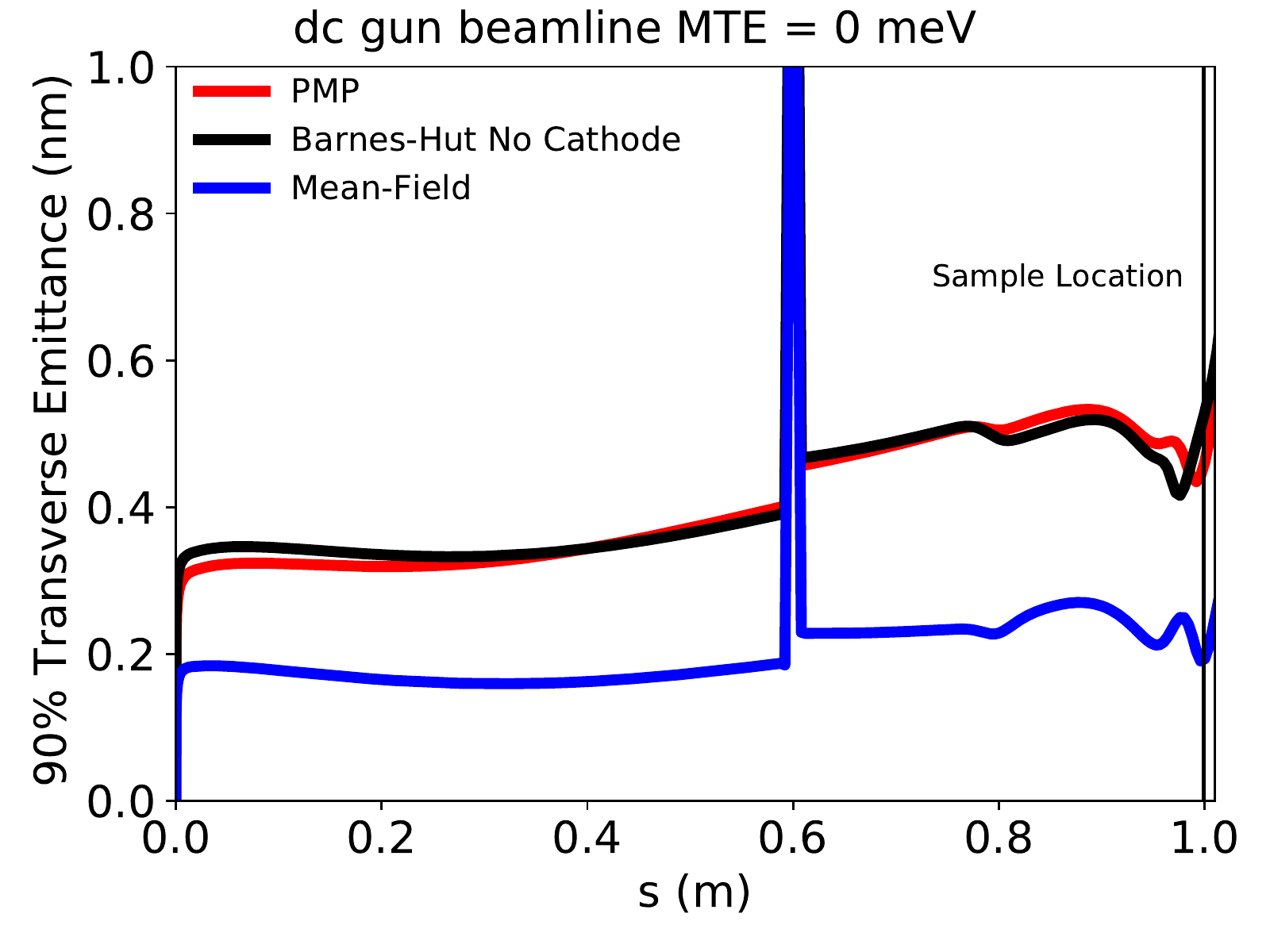}
\caption{}
\label{fig:NC_CdcE}
\end{subfigure}
\caption{Spot size and transverse normalized rms emittance comparison between the PMP method, Barnes-Hut method without a cathode, and mean-field space charge simulations of the dc UED beamline with 0 meV MTE.}  
\end{figure}

\begin{figure}

\begin{subfigure}[htbp]{0.45\textwidth}
\centering
\includegraphics[width=\linewidth]{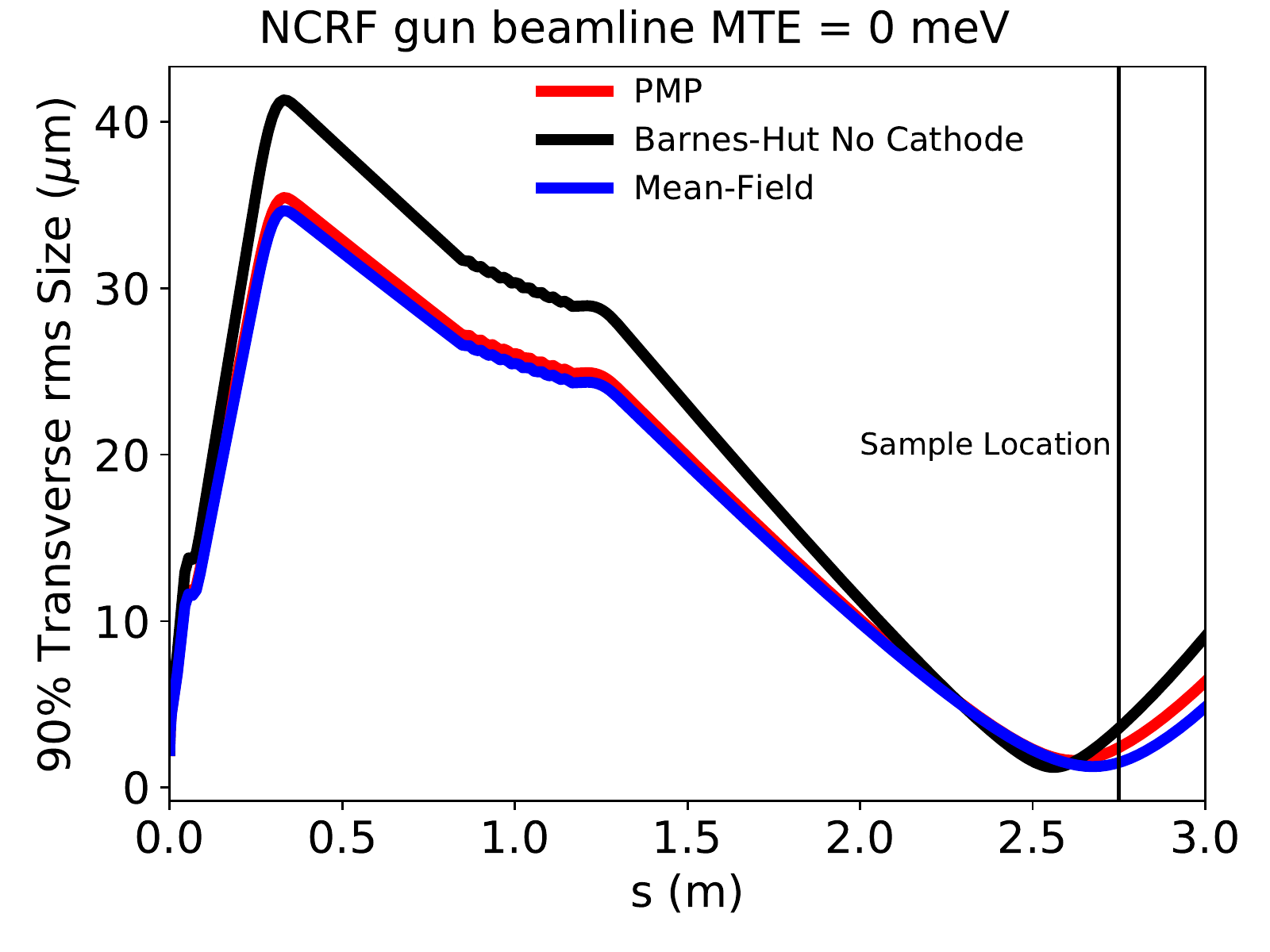}
\caption{}
\label{fig:NC_RFSS}
\end{subfigure}

\begin{subfigure}[htp]{0.45\textwidth}
\centering
\includegraphics[width=\linewidth]{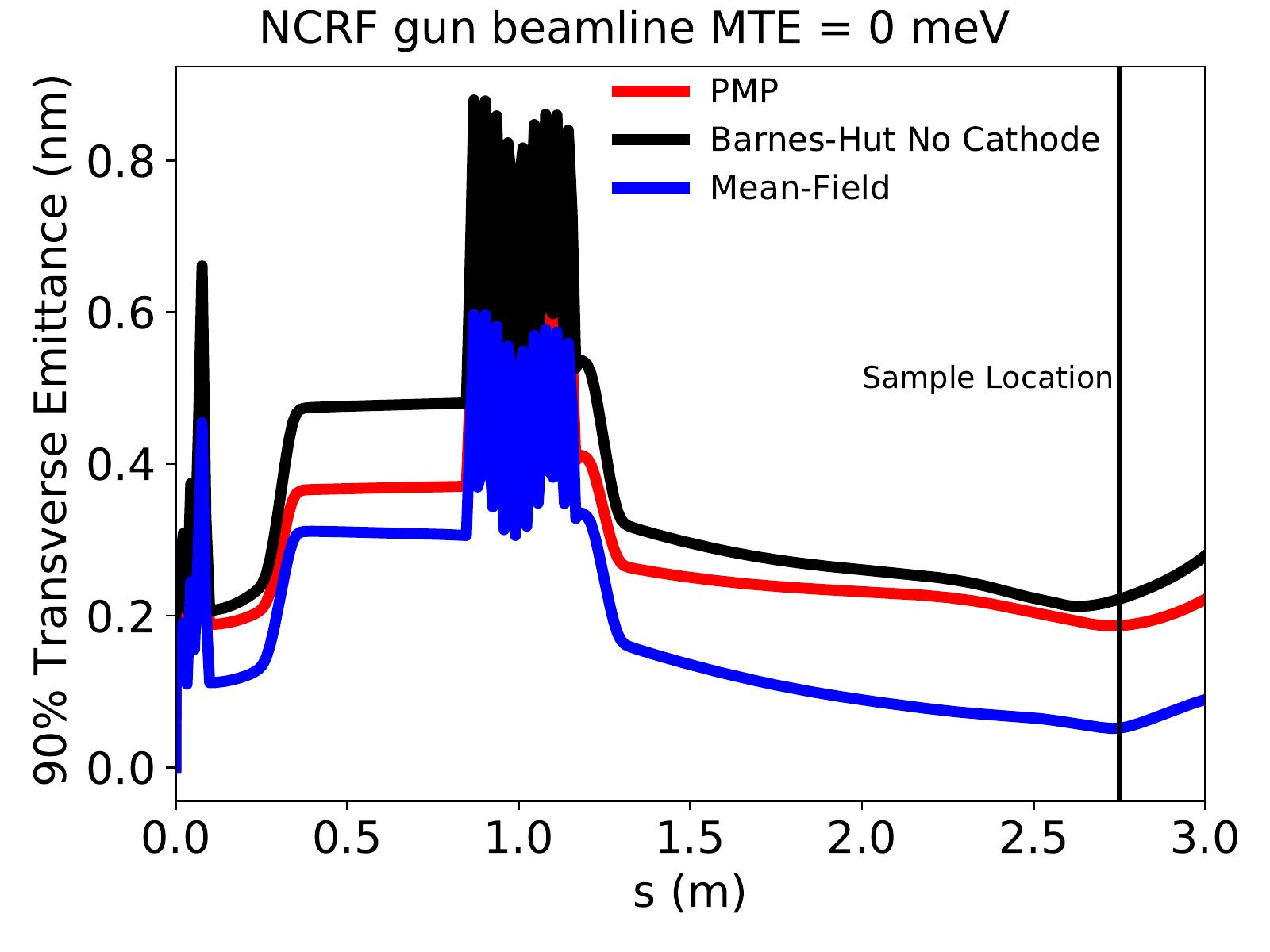}
\caption{}
\label{fig:NC_RFE}
\end{subfigure}
\caption{Spot size and transverse normalized rms emittance comparison between the PMP method, Barnes-Hut method without a cathode, and mean-field space charge simulations of the NCRF UED beamline with 0 meV MTE.}  
\end{figure}

The 90\% rms transverse  size of the beam along the dc and NCRF UED beamline is shown in Figs. \ref{fig:NC_CdcSS} and \ref{fig:NC_RFSS} respectively.  In all simulations, space charge increases the beam size, after which the first solenoid matches the beam size into the buncher cavity (which has a noticeable transverse defocusing), and the second solenoid forms the final waist. As expected from the emittance compensation process, the emittance minimum occurs very near the beam size waist.   

In the Barnes-Hut simulation without the cathode boundary condition, the initial space charge blowup leads to a larger spot size at the first solenoid.  Because of this, the focusing elements cause the beam waist to occur earlier than when the image force is included. Thus, omission of the image force generates significantly different beam dynamics.

The difference between the mean-field and PMP simulation is solely due to point-to-point effects, as a mean-field image force is included in each. It is interesting to note that the slightly larger spot size in PMP simulations translates to noticeably stronger focusing downstream. In the DC gun beamline, the stronger focusing is noted by a smaller beam size at the focus, and in the NCRF beamline, the beam waist is formed earlier by 5.5 cm. 

The evolution of the normalized transverse rms emittance for the dc and NCRF UED beamline with 0 meV MTE is shown in Figs. \ref{fig:NC_CdcE} and \ref{fig:NC_RFE} respectively.  Starting at an emittance of zero, the emittance quickly grows as the beams experience both the mean-field space charge phase space shearing, and also a growth in temperature due to the thermalization of the initial stochastic potential energy stored between near neighbors.

Comparing the transverse 90\% rms emittance of the Barnes-Hut method without a cathode to the PMP method, we find that qualitatively they behave similarly.  However, due to the earlier location of the beam waist in the Barnes-Hut simulations, the location of the emittance minima also shifts to an earlier position.

Although at a higher beam energy, we note the NCRF beamline exhibits a larger relative effect from point-to-point interactions. Later we will show that this can be explained by the effects of DIH with a larger initial electron number density. Though the space charge forces are more heavily suppressed at high energy, the effects of DIH thermalization occur at low energy near the cathode where relativistic suppression is negligible. The timescale of the evolution of the thermalization is a constant fraction of the plasma period \cite{Jared_DIH}, which in this case is roughly 30 ps. The value of the emittance is over a factor of 3.7 larger than when using only the mean-field approximation when including point-to-point effects.

It should be noted that because these beamlines were optimized to minimize the emittance of the mean-field space charge beam at the sample, the emittance in the PMP simulation is not optimized to be maximally compensated at its respective minimum.  Thus, these numbers represent an upper bound to the maximal effect of point-to-point space charge on the emittance for these UED beamlines. However, as we will show later, most of the emittance growth above the mean-field case  arises from microscopic DIH-like effects, which are insensitive to small perturbations in the focusing optics.



In all simulations, the bunch length is approximately constant until the buncher, after which it decreases linearly in time.  In the dc beamline, the bunch length at the sample is 0.92 ps in the PMP simulation and 0.91 ps in the mean-field simulation.  In the NCRF beamline, the bunch length at the sample is 0.91 ps in both simulations. Thus, point-to-point effects do not play a large role in determining bunch length at the sample.

\section{Microscopic Evolution}
We now move to analyze the microscopic real and phase space distributions in an effort to determine to what extent they follow simple predictions of beam heating via DIH, and to what extent this heating determines the total rms emittance growth from point-to-point effects.

\subsection{Core Emittance}

\begin{figure}[htbp]
\centering
\includegraphics[width=220pt]{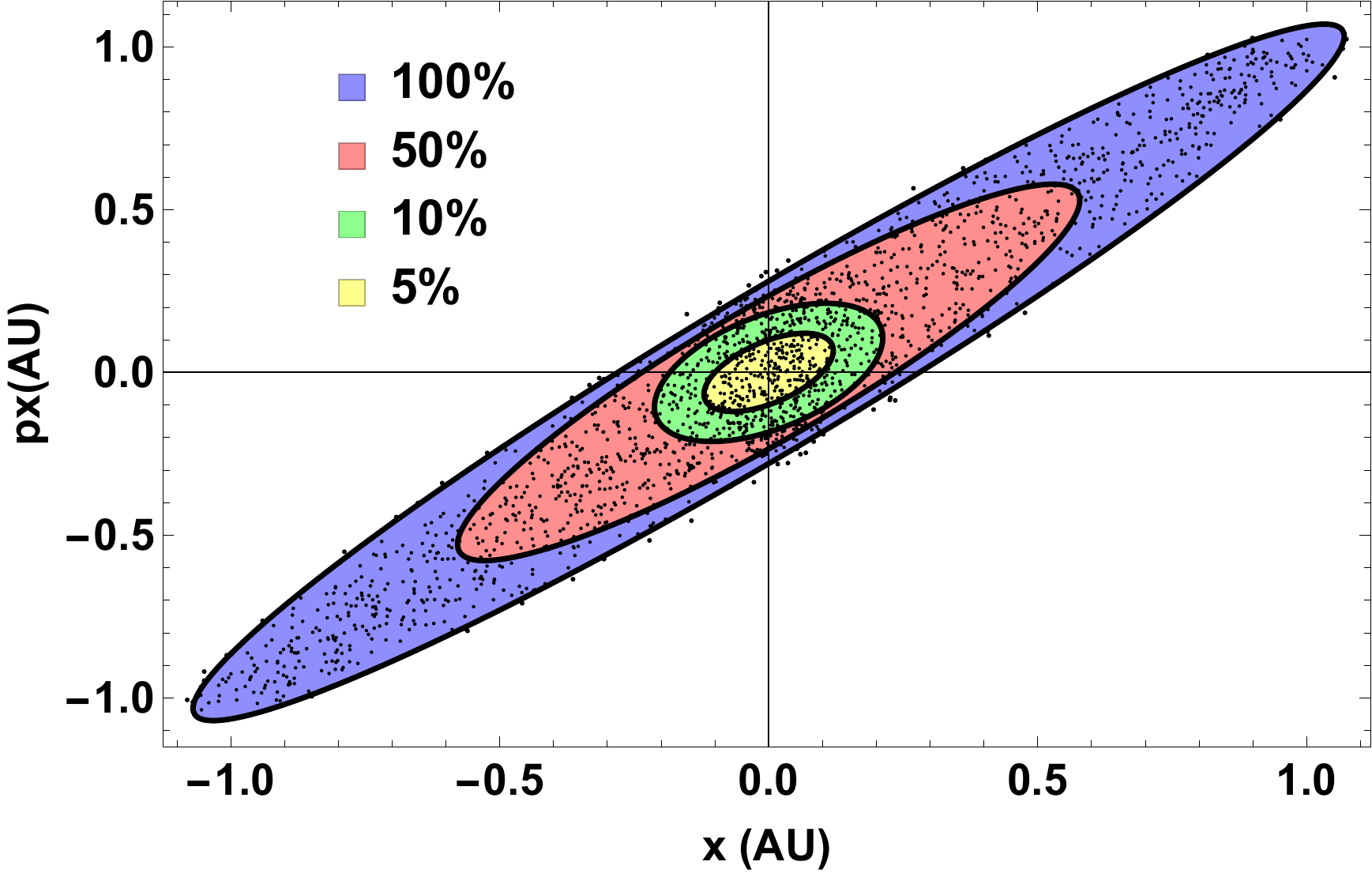}
\caption{Depiction of emittance vs. particle fraction selection.  Ellipses are drawn such that they represent the phase space area occupied by the beam using only a given fraction of the total number of particles.  Ellipse dimensions are selected such that the emittance is minimized for each particle fraction.}  
\label{fig:PF}
\end{figure}

\begin{figure*}
    \centering
    \begin{subfigure}[t]{0.45\textwidth}
        \centering
        \includegraphics[width=\linewidth]{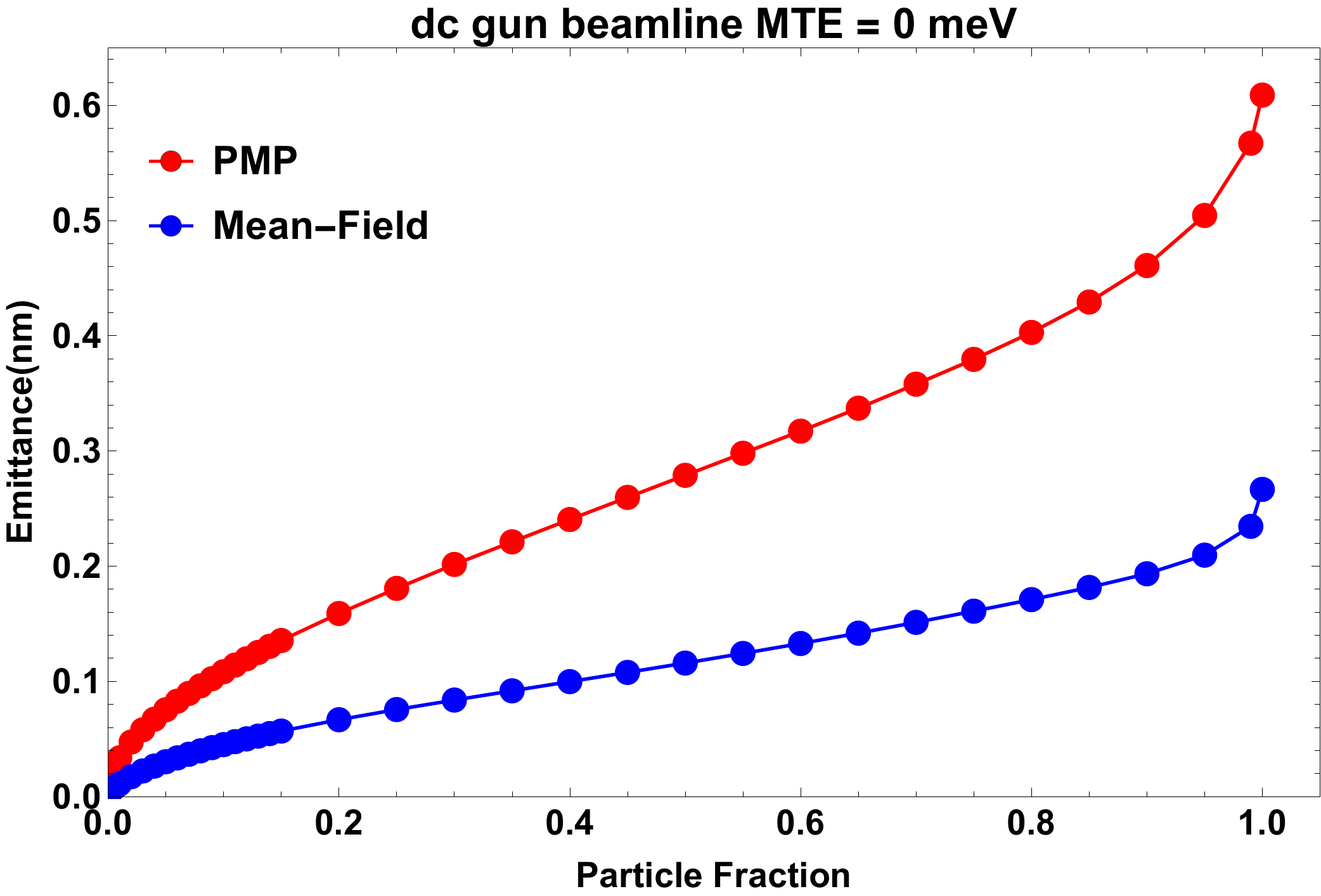}
        \caption{}
        \label{fig:CdcPF}
    \end{subfigure}%
    ~
    \begin{subfigure}[t]{0.45\textwidth}
        \centering
        \includegraphics[width=\linewidth]{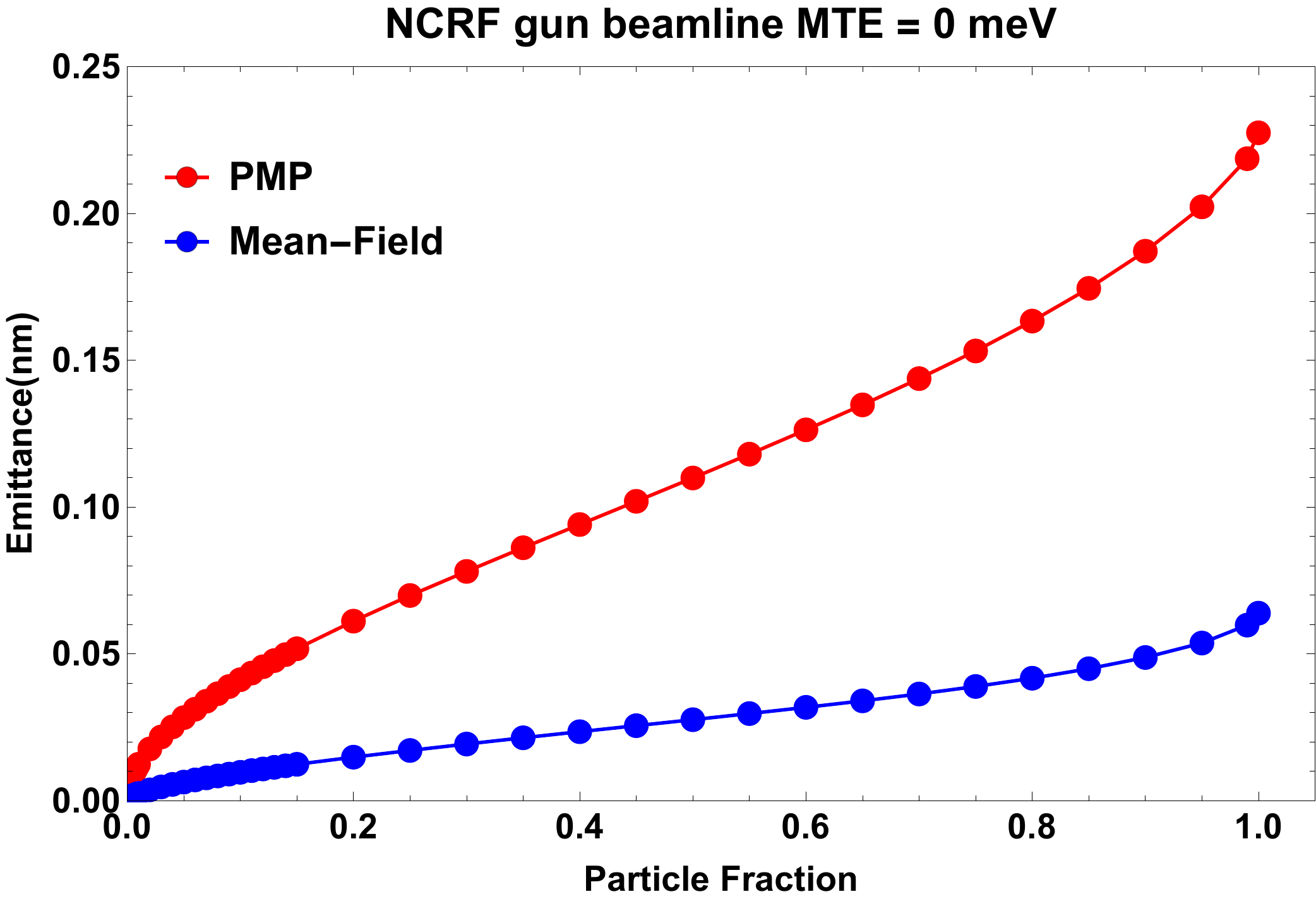}
        \caption{}
        \label{fig:CRFPF}
    \end{subfigure}
    
    \begin{subfigure}[t]{0.48\textwidth}
    \centering
    \includegraphics[width=\linewidth]{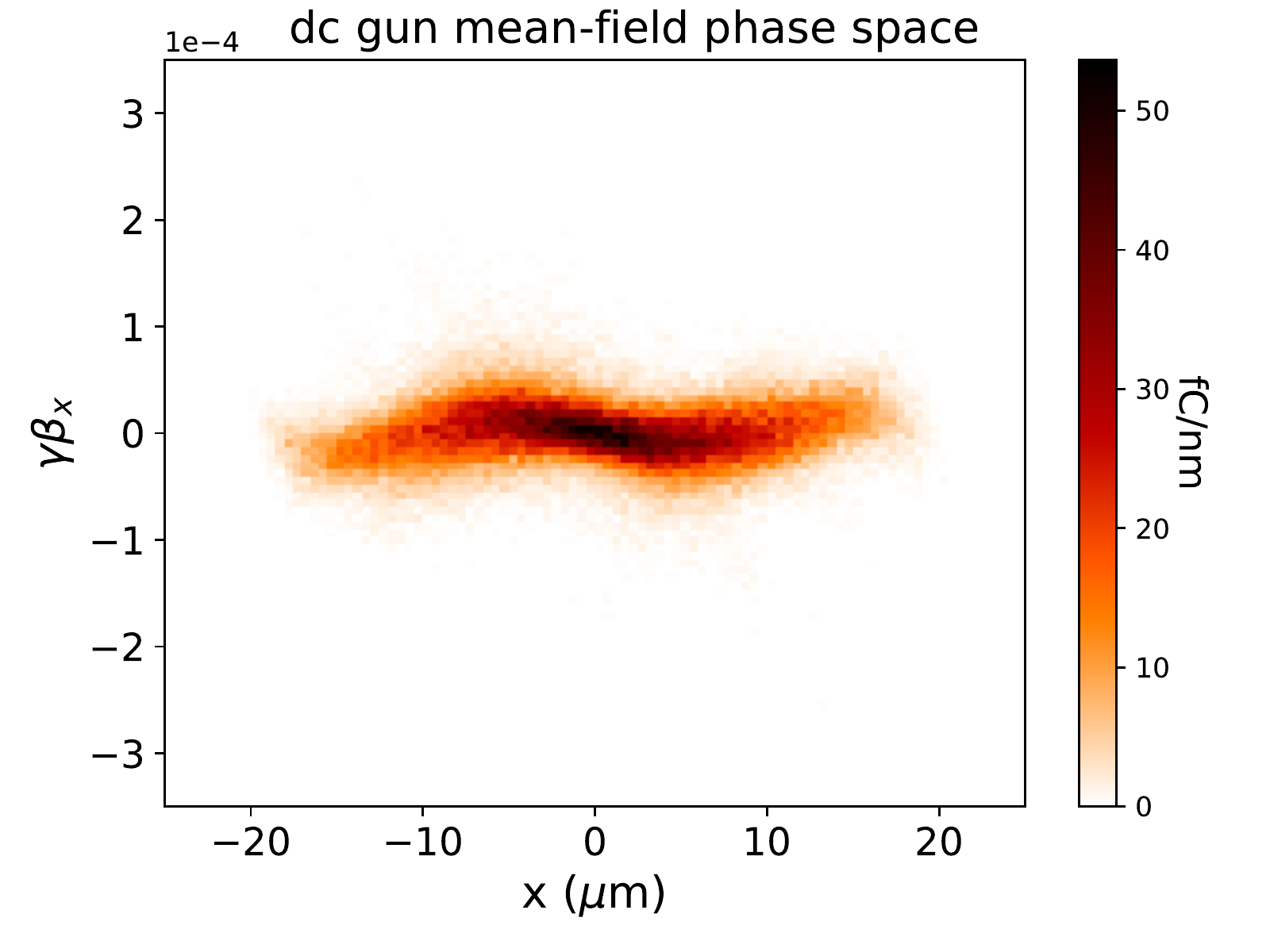}
    \caption{}
    \label{fig:DCMFPS}
    \end{subfigure}%
    ~
    \begin{subfigure}[t]{0.48\textwidth}
        \centering
        \includegraphics[width=\linewidth]{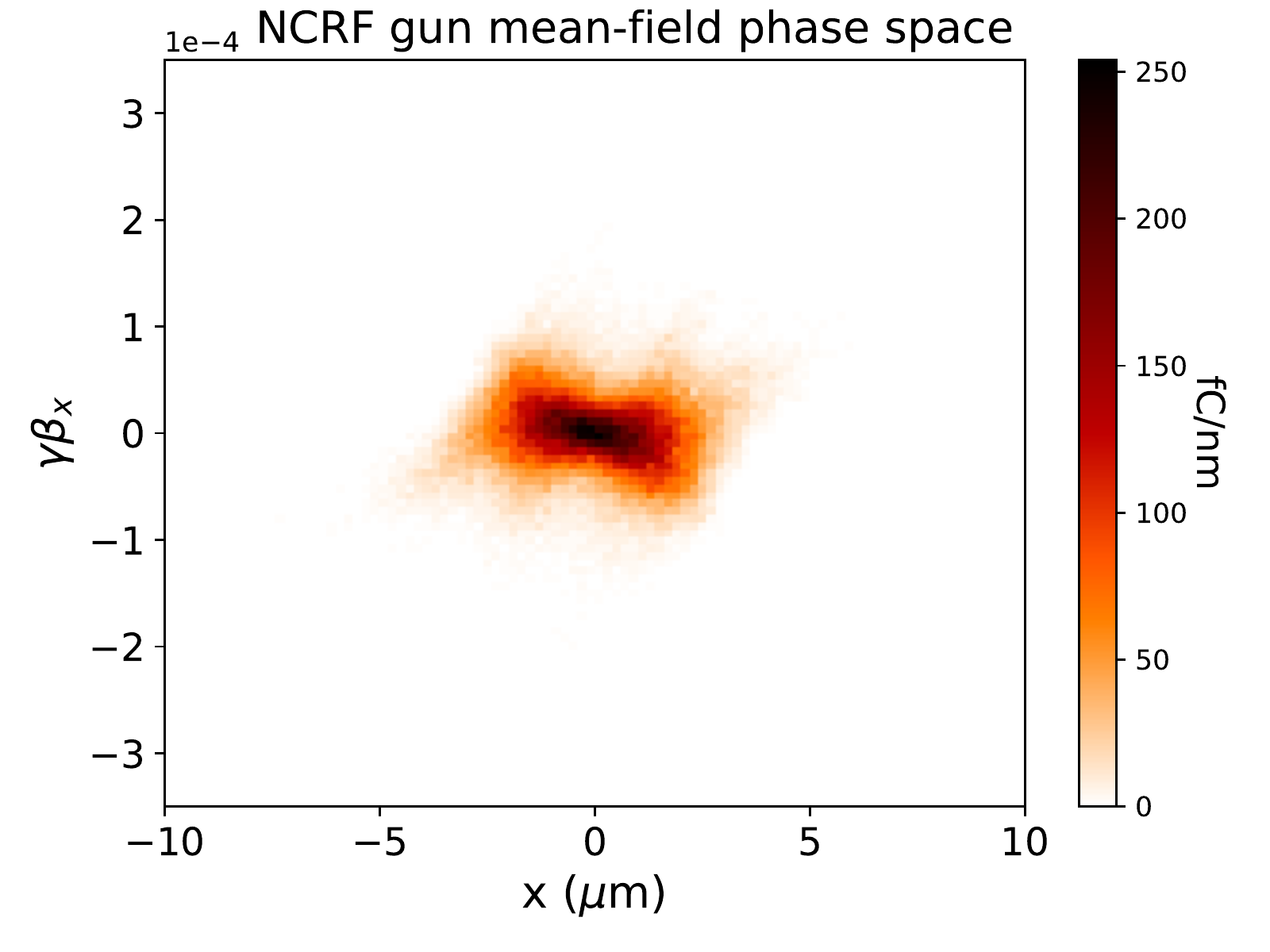}
        \caption{}
        \label{fig:RFMFPS}
    \end{subfigure}
        
    \begin{subfigure}[t]{0.48\textwidth}
    \centering
    \includegraphics[width=\linewidth]{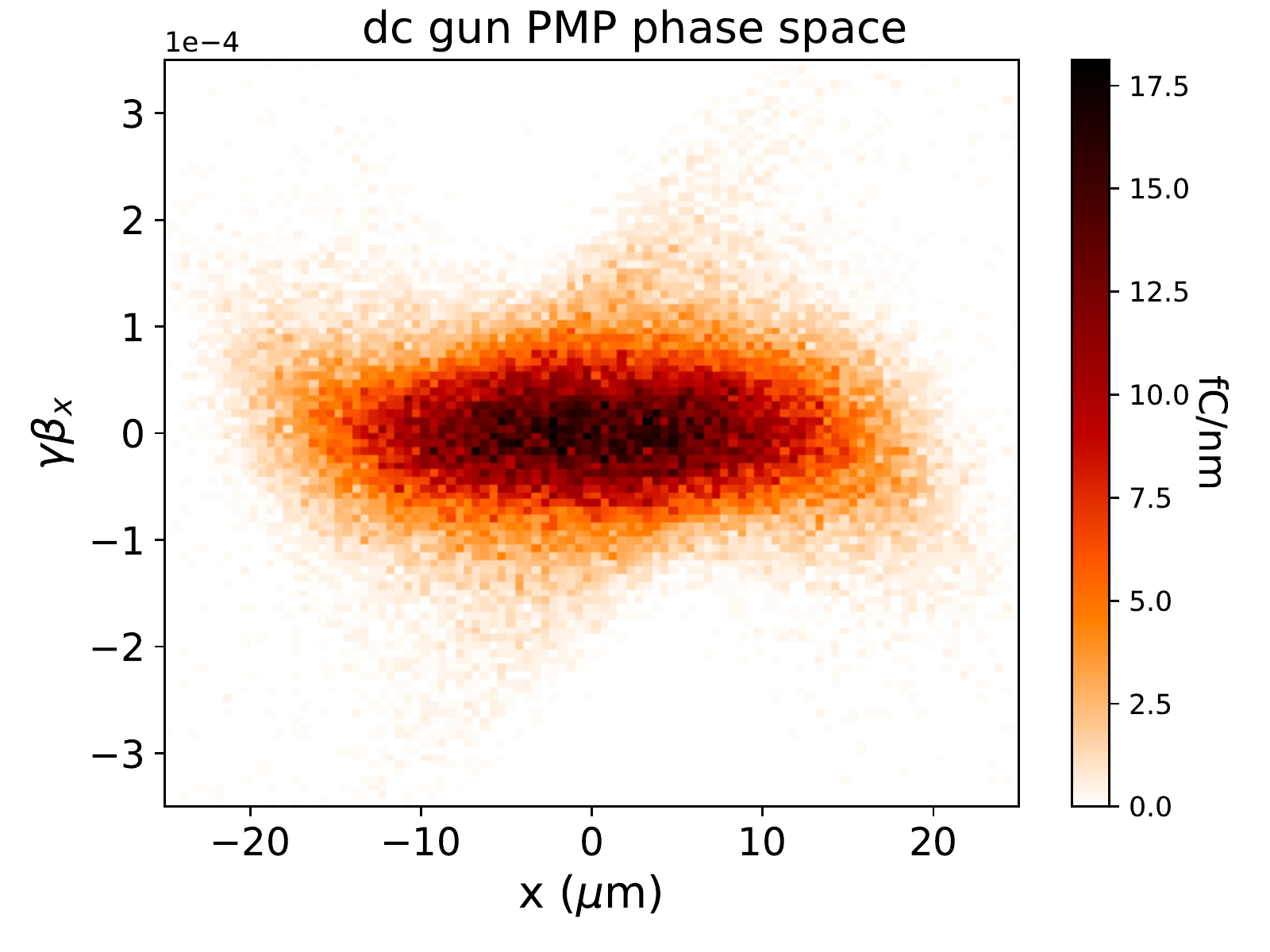}
    \caption{}
    \label{fig:DCPMPPS}
    \end{subfigure}%
    ~
    \begin{subfigure}[t]{0.48\textwidth}
        \centering
        \includegraphics[width=\linewidth]{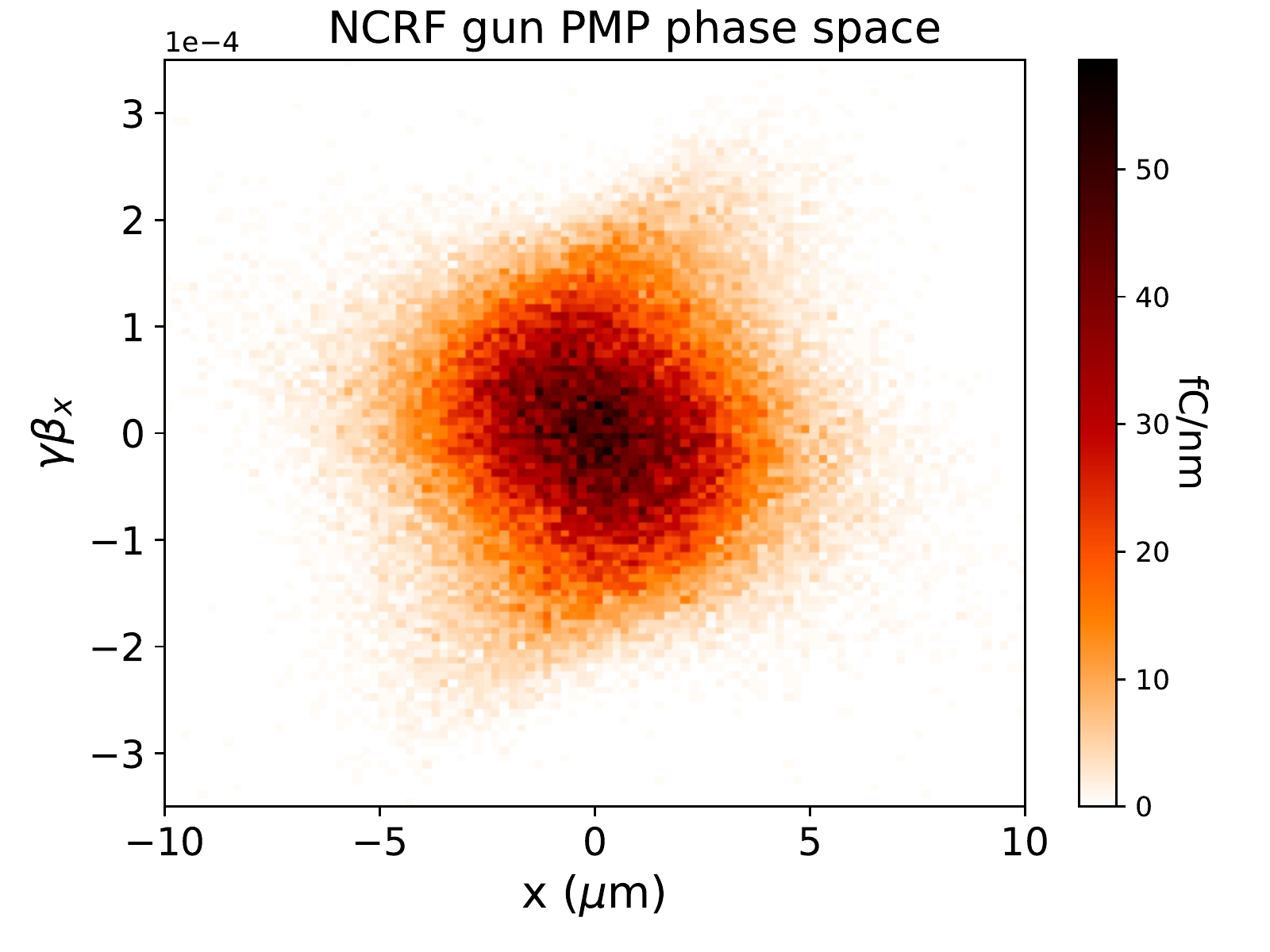}
        \caption{}
        \label{fig:RFPMPPS}
    \end{subfigure}
    \caption{Transverse normalized rms emittance vs. particle fraction and phase space comparison between PMP and mean-field simulations of the 2 UED beamlines at the respective emittance minimum near the end of the beamline for each simulation.    Phase space portraits are shown with linear $x - p_x$ correlation removed.}
\end{figure*}

One tool which can be used to analyze the microscopic evolution of a beam is the core emittance. The core emittance is a measure of the peak transverse phase space density.  It is defined through an emittance vs. particle fraction curve  \cite{CE_Main}.  Starting with the full beam emittance (particle fraction of 1), particles can be excluded from the emittance calculation such that the resulting emittance is minimized, see Fig. \ref{fig:PF}.  The core emittance is defined as the limit of the slope of the emittance vs. particle fraction curve as the particle fraction goes to 0. It is inversely proportional to the peak transverse phase space density:

\begin{equation}
\epsilon_{c}=\frac{d\epsilon}{df} \Big |_{f\rightarrow 0^+}=\frac{1}{4\pi \rho_0}
\label{ce_eq}
\end{equation}

\noindent where $\epsilon$ is the emittance of the beam for a given fraction of particles, $f$ is the particle fraction and $\rho_0$ is the peak phase space density. We expect $\rho_0$ to be invariant in mean-field space charge systems, but the introduction of point-to-point effects can break this invariance. However, because we compute the core emittance with a finite number of beam particles and a finite number of bounding ellipses the value will never be exactly zero even if we start with zero MTE.

In Figs. \ref{fig:CdcPF} and \ref{fig:CRFPF}, the emittance vs. fraction curves at the emittance minima are shown for mean-field and point-to-point space charge for the dc gun and NCRF gun UED beamlines. Note the sharp increase in rms emittance due to outliers when the particle fraction approaches unity.  As can be seen, for small particle fractions the emittance of the PMP simulation is significantly higher than that of the mean-field simulation.  This shows that point-to-point effects not only have created more outliers, but have fundamentally degraded beam quality up to and including the core of the beam, as would be expected from DIH.  To help illustrate this further, phase space portraits are shown at the respective emittance minima for the two beamlines with mean-field space charge in Figs. \ref{fig:DCMFPS} and \ref{fig:RFMFPS} and with PMP space charge in Figs. \ref{fig:DCPMPPS} and \ref{fig:RFPMPPS}. The decrease in core phase-space density is clearly seen by the increased width in the $\gamma \beta_x$ coordinate in Figs. \ref{fig:DCPMPPS} and \ref{fig:RFPMPPS}. Note that the faint diagonal tails in these figures are not outliers due to stochastic interactions, but the effect of slightly mistuned optics.

\begin{figure}[htbp]
\centering
\includegraphics[width=220pt]{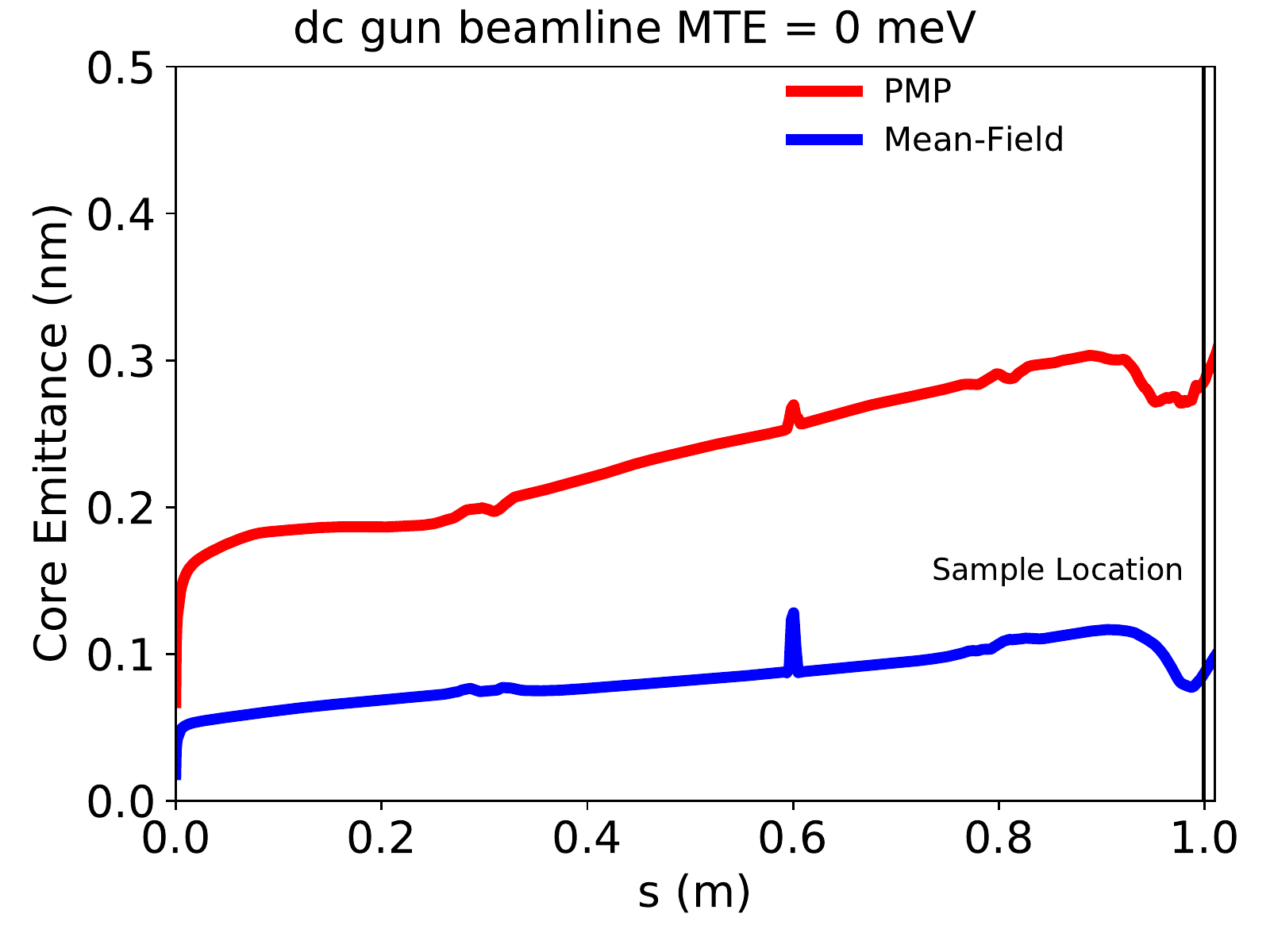}
\caption{Core emittance comparison between PMP and mean-field simulations of the dc UED beamline with 0 meV MTE.}
\label{fig:CdcCE}
\end{figure}

\begin{figure}[htbp]
\centering
\includegraphics[width=220pt]{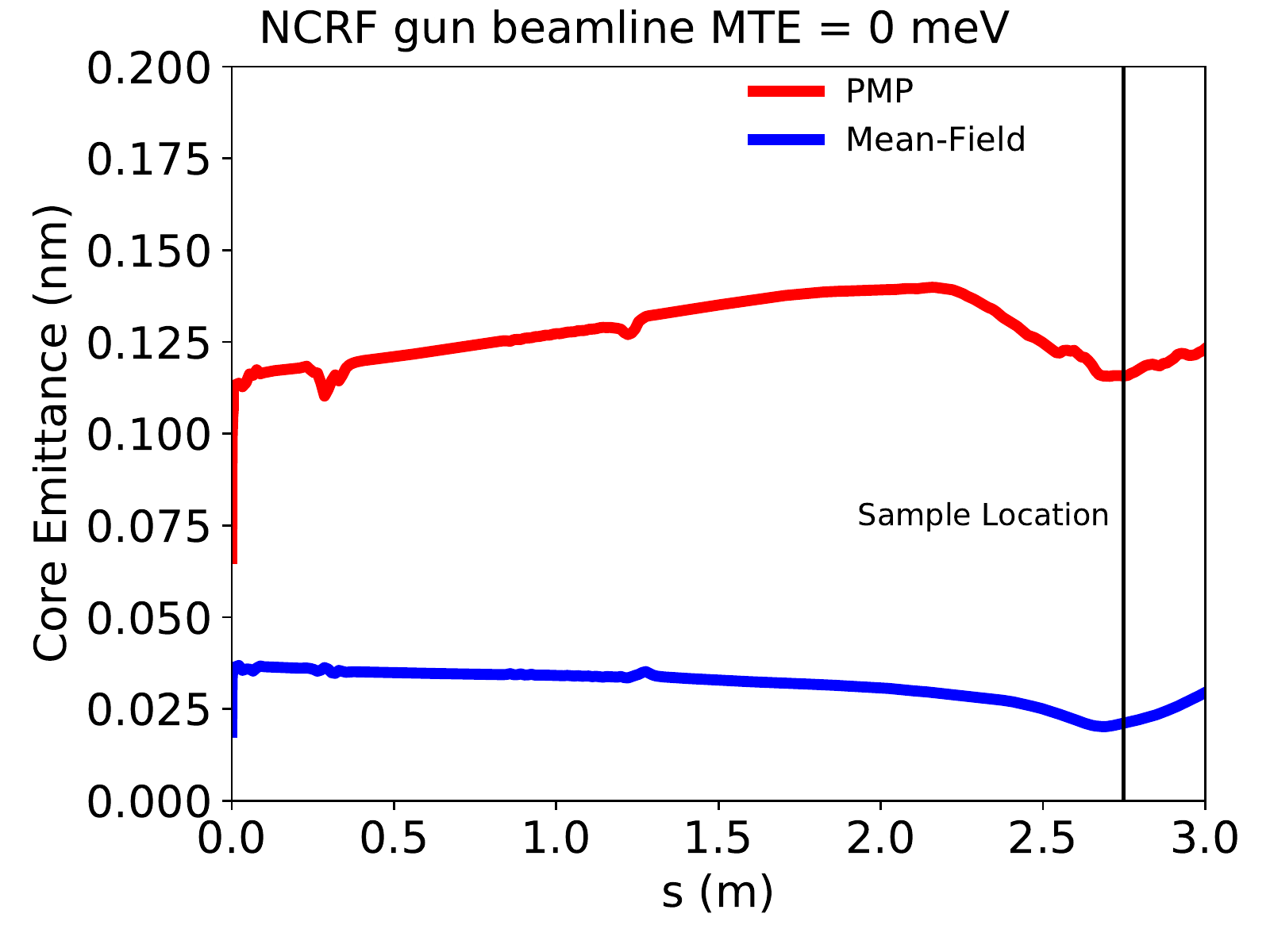}
\caption{Core emittance comparison between point-to-point and mean-field space charge for the NCRF UED beamline with 0 MTE.}
\label{fig:CRFCE}
\end{figure}

The core emittance of the beam was computed at different points along the dc beamline, see Fig. \ref{fig:CdcCE}, and Fig. \ref{fig:CRFCE} for the NCRF gun beamline.  After a quick initial rise  at low energy, the core emittance in the point-to-point simulations remains far above that of the mean-field simulation. 

The core emittance at the sample of these simulations is shown in table \ref{tab:CE}. As with the transverse rms emittance, the effects of point-to-point space charge are more distinct in the NCRF beamline, where the number density of electrons is higher. 

\begin{table}[htb]
\caption{Core emittance with 0 meV MTE at sample}\label{tab:CE}
\begin{tabular}{||c| c | c||} 
 \hline
 Beamline & PMP $\epsilon_c$ (nm) & mean-field $\epsilon_c$ (nm) \\ [0.5ex] 
 \hline
 dc & 0.28 & 0.08 \\ 
 \hline
 NCRF & 0.12 & 0.020  \\
 \hline
\end{tabular}
\end{table}

\subsection{Radial Distribution Function}
The radial distribution function, $g(r)$, of a system of particles relates the bulk density of particles to the local particle density as a function of distance from a reference particle \cite{RDF}.  A microscopically uniform distribution of particles has a constant radial distribution function excluding effects from finite system size. In this case, a neighbor particle to a given reference particle is equally likely to be found at any distance.   However, due to the divergence of the Coulomb interaction, the number of very near neighbors to a reference particle in an electron beam evolves to become zero. This is known as the Coulomb hole and it results in a decrease of the total potential energy of the system \cite{C_Hole}. This release of potential energy causes disorder induced heating, and we denote the resulting mean kinetic energy each particle gains from this heating by $E_{\rm DIH}$.

\begin{figure}[htp]
\centering
\includegraphics[width=220pt]{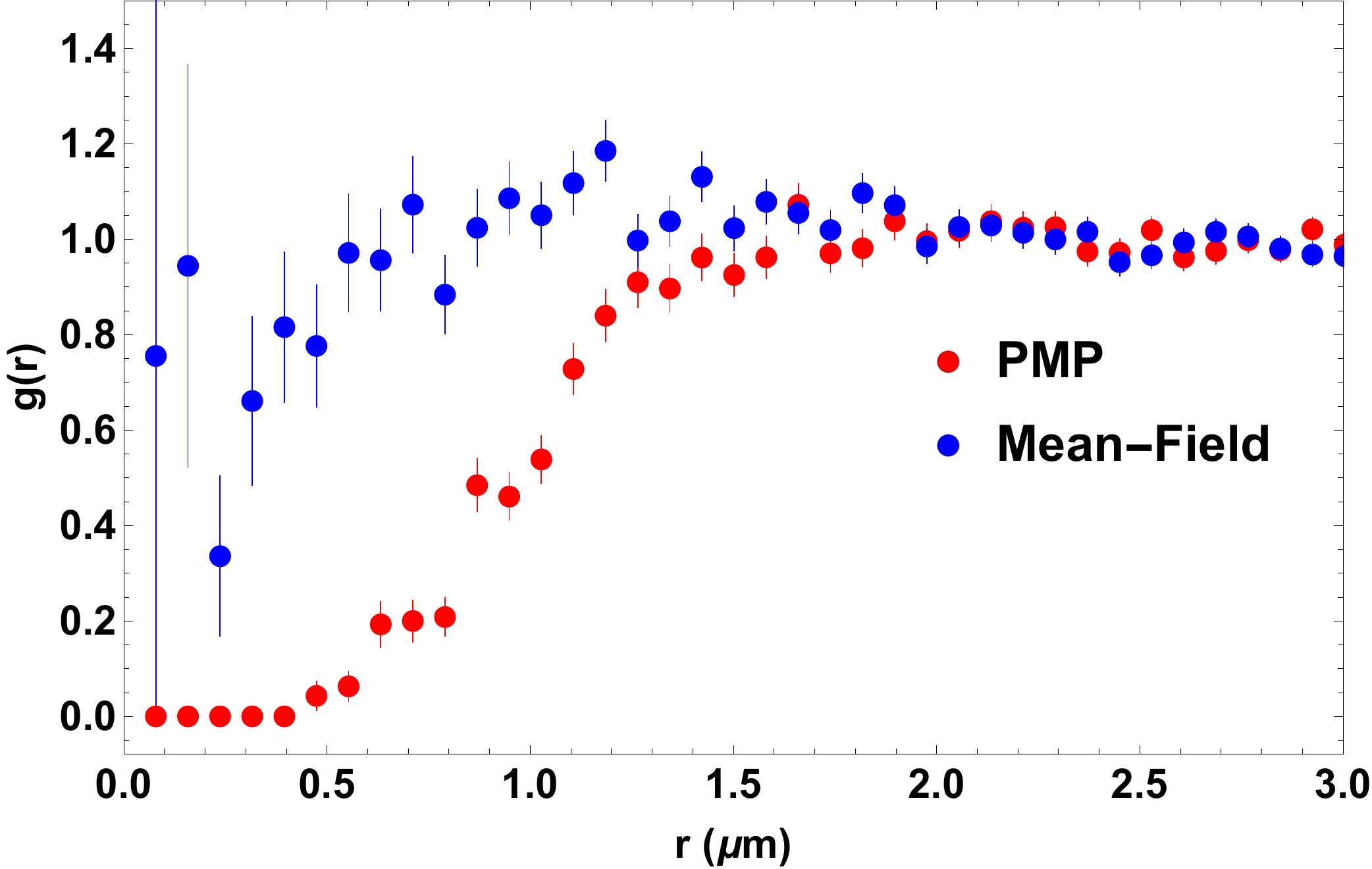}
\caption{Radial distribution function comparison between PMP and mean-field simulations of the NCRF UED beamline with 0 meV MTE $\sim 3$mm away from the cathode.  Only a small $r$ portion of the distribution is plotted to show the creation of the Coulomb hole when point-to-point space charge is used.  For comparison, the distributions were normalized such that the mean of the radial distribution functions from 1.5$\mu$m to 3.0$\mu$m is equal to 1.}
\label{fig:grRF}
\end{figure}

To calculate $g(r)$, the following procedure is used:  Screen based outputs are taken from a chosen position along the beamline.  Because we are interested in the dynamics of the core of the beam, only the 10\% of electrons which are closest to the longitudinal center of the beam are used. For each of these particles, the distance to every other selected particle is calculated.  By making a histogram of these distances, we generate a plot of $g(r) \times \rho N \times 4 \pi r^2 \Delta r$ as a function of $r$ , where $N$ is the number of particles, $\rho$ is the bulk volume density of those particles, and $\Delta r$ is the bin size.  A statistical uncertainty is assigned to each bin equal to the square root of the number of particles in the bin.  Dividing out the $r^2$ term, $g(r)$ is found up to numerical prefactors.

The radial distribution function shortly after emission in the case of the NCRF UED beamline with 0 meV MTE is shown in Fig. \ref{fig:grRF}.  For distances smaller than 1.5 $\mu$m, the radial distribution function in the case of point-to-point space charge decreases to 0 as expected, while $g(r)$ of mean-field space charge does not.  This same behavior can be seen for the dc beamline with an MTE of 0 meV.

\subsection{Disorder Induced Heating Calculation}

From the radial distribution function, the potential energy of a particle due to its surrounding particles can be calculated as:
\begin{equation}
E_{\rm potential}=\int_0^\infty  4 \pi r^2 \rho g(r) u(r) \mathrm{d}r
\label{eq:Ep}
\end{equation}
where $\rho$ is the bulk volume density of the beam and $u(r)$ is the potential energy of two electrons particles separated by a distance $r$ \cite{E_DIH_Ref}. Using equation \ref{eq:Ep}, $E_{\rm DIH}$ can be calculated by finding the difference between  the potential energy calculated via the radial distribution function in the point-to-point simulation and a calculation using the same radial distribution function where the  the Coulomb hole is artificially filled. Tests with stationary distributions which have known $E_{\text{DIH}}$ show that this estimation method is accurate to within 20\%, with discrepancies arising primarily due to the determination of the peak location of $g(r)$.

This energy difference would be $E_{\rm DIH}$ for all times after heating if the beam did not change in size throughout the simulation.  Because the beam size changes, additional calculation is required to recover $E_{\rm DIH}$.  If the beam changes in a self-similar way, such that its aspect ratio remains constant, the energy found through this subtraction is $E_{\rm DIH}$ multiplied by the ratio of the initial average interparticle distance (defined below) to the current average interparticle distance.  This can be seen by investigating the radial scaling of equation \ref{eq:Ep} noting that $\rho \propto 1/r^3$ and $u(r) \propto 1/r$.  Thus by multiplying by the inverse of this factor, we can estimate  $E_{\rm DIH}$  assuming DIH takes place very near the cathode. Further, it is clear the assumption of self-similarity is invalid if the beam deviates significantly from its initial aspect ratio, which can occur when space charge forces cause significant ``blowout" (spatial \cite{Cigar} or longitudinal \cite{MusPRL}) and near most beam waists.

The initial interparticle spacing requires definition, as at $t=0$, no beam yet exists.  To do this, we will approximate the beam as a uniform cylinder with equivalent rms sizes as at the cathode surface.  Assuming a uniform acceleration over the small length and time scale the beam is being emitted from the cathode, the front of the beam will travel to a distance of $L=\frac{1}{2} a_{E_0} t^2$, where $a_{E_0}$ is the acceleration of an electron in a uniform electric field $E_0$, and $t$ is the difference in time between the first and last particle emitted.  Approximating as a uniform distribution where $R= 2\sigma_R$ and $t = \sqrt{12}\sigma_t$, the volume of the beam can be found as:
\begin{equation}
    V = \pi R^2 L
      \approx \frac{24 \pi e E_0}{mc^2}\sigma_x^2(c\sigma_t)^2
\end{equation}

Using this volume, an initial average interparticle distance can be found as $\rho \approx (V/N)^{1/3}$.

\begin{figure}[htp]
\centering
\includegraphics[width=220pt]{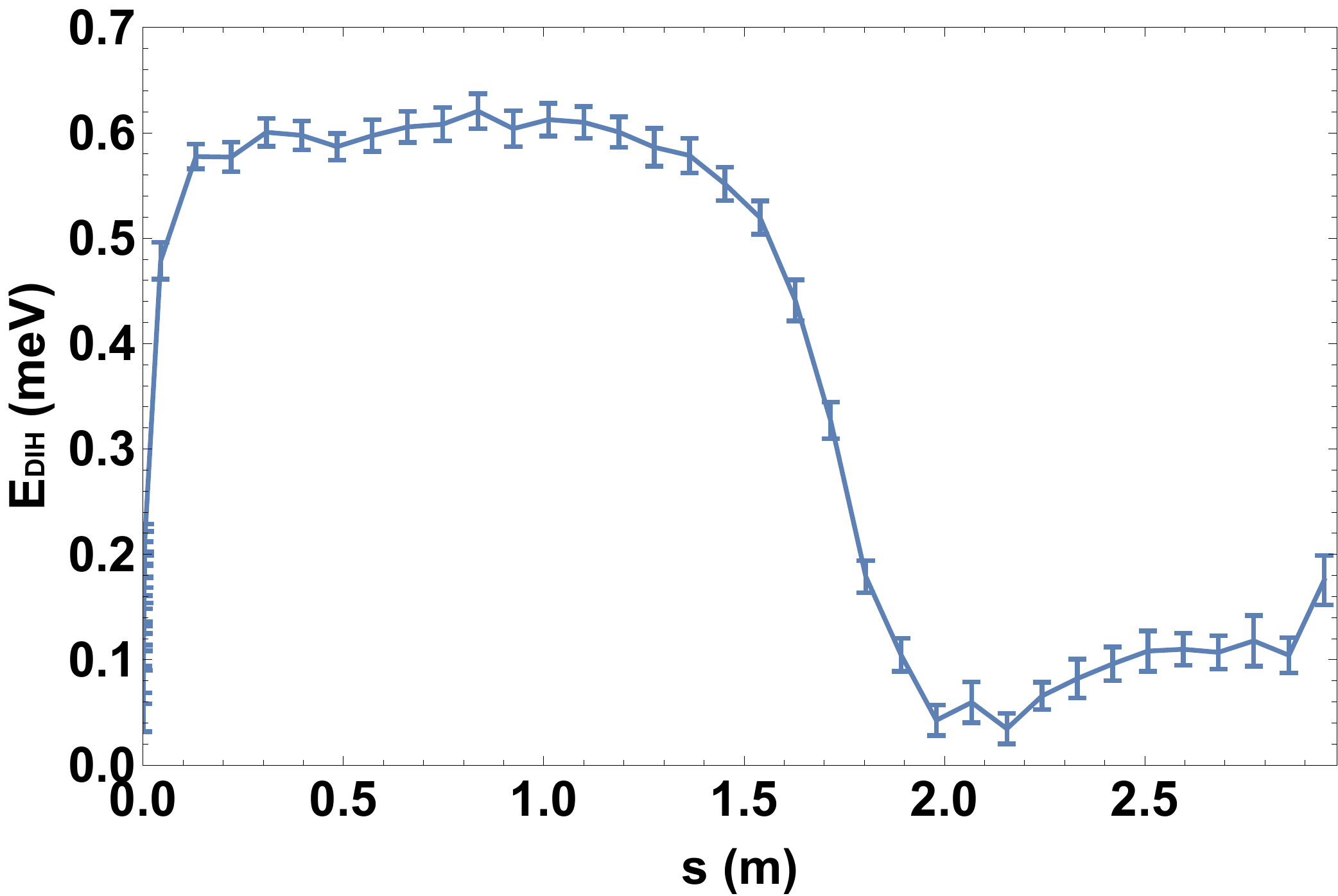}
\caption{Energy from disorder induced heating as calculated from $g(r)$ in the NCRF beamline with 0 MTE and an initial density of $10^{17}\text{ m}^{-3}$.}
\label{fig:DIH_8}
\end{figure}

\begin{figure}[htp]
\centering
\includegraphics[width=220pt]{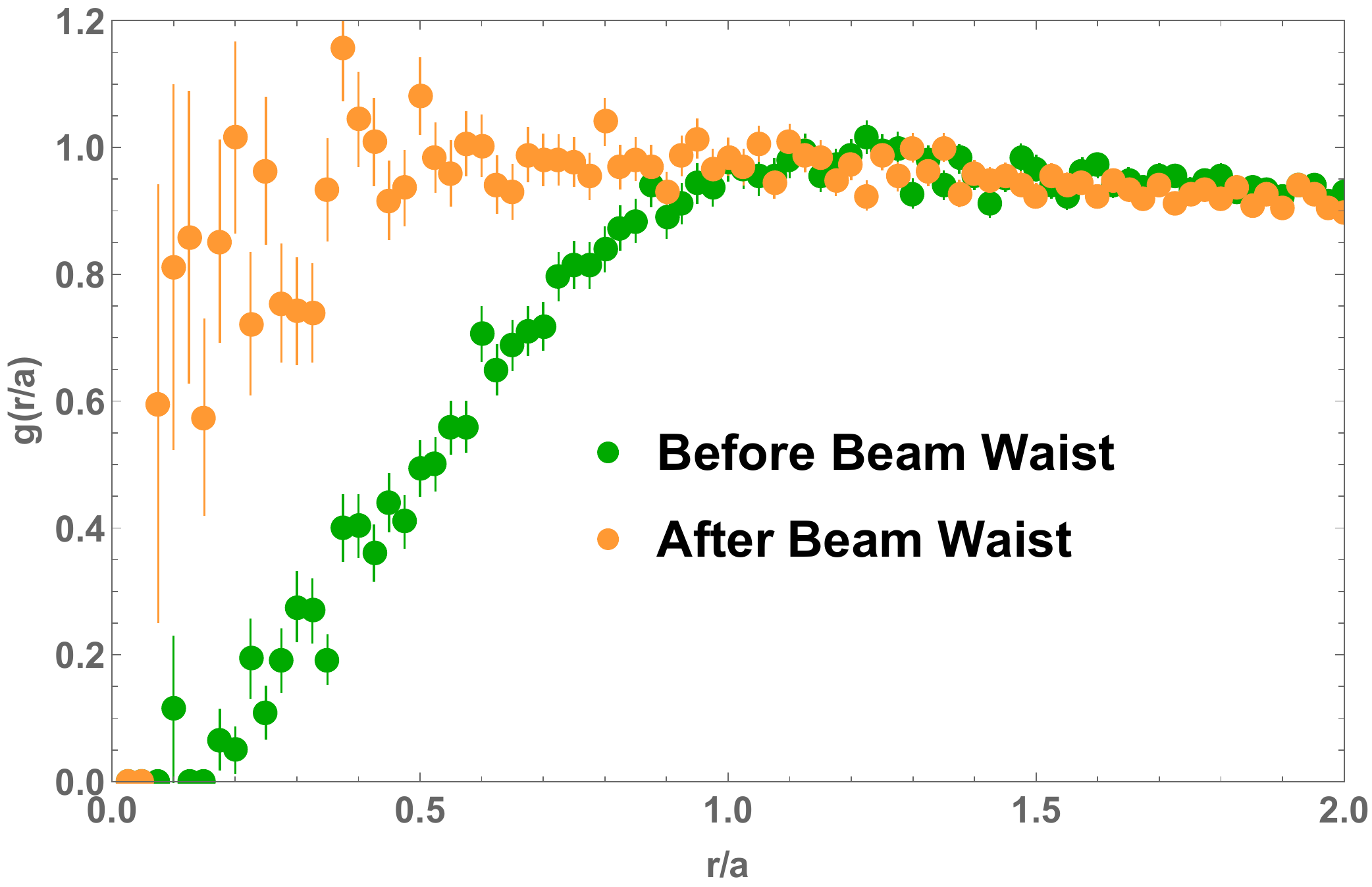}
\caption{Radial distribution function comparison between of the NCRF beamline with an MTE of 0 meV and an initial density of $10^{17}\text{ m}^{-3}$ before and after the beam waist.  For comparison, the distances were normalized by the average interparticle distance, $a$, and the radial distribution functions, $g(r/a)$, were normalized such that $g(r/a=1.25) = 1$.}
\label{fig:Cool}
\end{figure}

A plot of the $E_{\rm DIH}$ estimate is shown in Fig. \ref{fig:DIH_8}.  There are three main features of this plot. First there is an initial rise in $E_{\rm DIH}$ corresponding to the time it takes for the Coulomb hole to form, i.e. the inverse plasma oscillation frequency of the beam:
\begin{equation}
\tau \approx 0.3 \frac{2\pi}{\omega_p} =  0.6 \pi\left(\frac{n_0 e^2}{m \epsilon_0}\right)^{-1/2}
\end{equation}
where $n_0$ is the density of the electron beam \cite{Jared_DIH}.  After the initial rise there is a plateau.  The mean value of this plateau is used for the value of $E_{\rm DIH}$ of the simulation, and the standard deviation of these values is treated as an uncertainty.  The drop in $E_{\rm DIH}$ corresponds to the transverse focus of the beam.  During this focusing, not only is the self-similarity assumption violated, there is another microscopic reorganization in which the Coulomb hole is filled. This is shown in Fig. \ref{fig:Cool}. Please note that this downstream filling in the Coulomb hole does not have a significant impact on the core emittance of the beam anymore. This is because near the beam waist, the transverse temperature of the of the beam  $mc^2(\epsilon/\sigma_x)^2$ is $\sim 12$ meV and the energy per particle required to fill the Coulomb hole is $\sim 0.3$ meV.

\subsection{Disorder Induced Heating Density Dependence}

\begin{figure}[htbp]
\centering
\includegraphics[width=220pt]{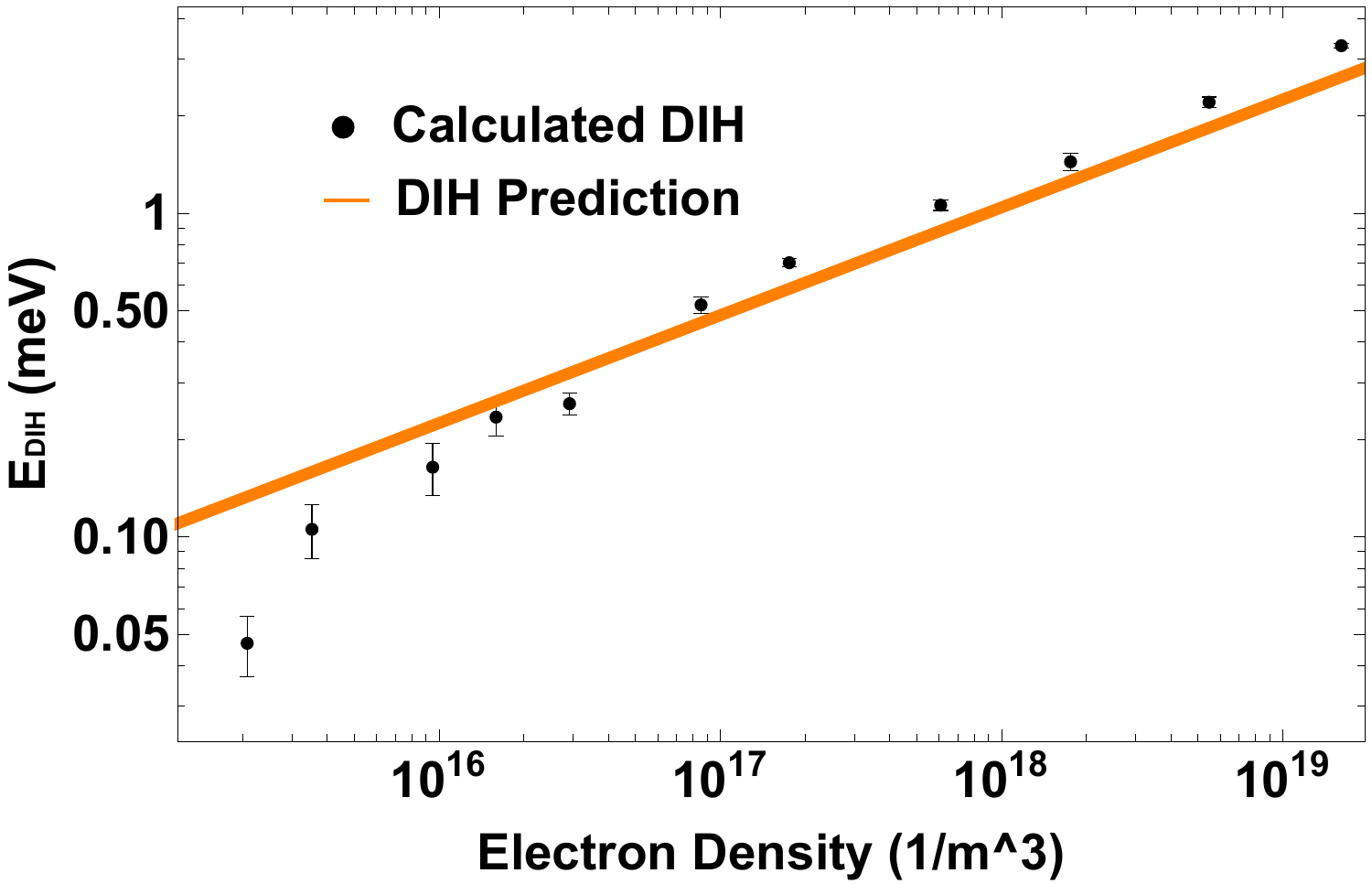}
\caption{Disorder induced heating as calculated from equation \ref{eq:DIH} compared to the result calculated from simulation.}
\label{fig:DIH_DD}
\end{figure}

For a stationary electron plasma with a starting temperature of zero, the energy released by disorder induced heating can be calculated via Eq. \ref{eq:DIH}. To test the density dependence of equation \ref{eq:DIH} in the realistic, non-stationary case, this procedure was repeated on simulations of the same NCRF beamline with 0 MTE, while changing the radius of the inital beam in order to alter its density.  The result is shown in Fig. \ref{fig:DIH_DD}.

For densities of $10^{16}$~$\text{m}^{-3}$ and above, the simulation results agree with the simple stationary theory within a factor of 2.  At a density near $10^{15} m^{-1/3}$, the timescale for heating is $\sim 1$ ns, which is approximately the time it takes the beam to enter the first solenoid.  Because the beam has had time to change significantly in size and shape, there is no reason to expect that the approximations used in calculating $E_{\rm DIH}$ remain true, thus it will be ignored in the following analysis. We find a that $E_{\rm DIH}$ in our simulations scales with density to the power $0.39\pm 0.02$, close the value of $1/3$ in Eq. \ref{eq:DIH}. 

\subsection{Core Emittance and rms Emittance Contributions from Disorder Induced Heating}
From the calculated values of $E_{\rm DIH}$, we can estimate the expected increase in the core emittance, and rms transverse emittance from DIH alone. This will help determine to what extent the Coulomb hole formation determines the growth in core and rms emittance.

For the core emittance, starting from equation \ref{ce_eq}, the density in x-$p_x$ space at the transverse origin can be calculated assuming a cylindrical beam shape and a Gaussian momentum distribution:

\begin{equation}
    \epsilon_c= \frac{\sqrt{2\pi}}{4}\sigma_x \sigma_{p_x}
\end{equation}

The initial spread of momenta can be written in terms of the MTE in the standard way:
\begin{equation}
\sigma_{p_x} = \sqrt{\frac{\text{MTE}}{mc^2}}
\end{equation}

For the presented simulations, the initial MTE is 0, but some of the disorder induced heating energy will be released in the transverse phase space, resulting in a non-zero momentum spread.  In general, the distribution of the heating will depend on the shape of the beam. However, in the case that the average inter-particle distance is much less than the smallest length scale of the beam, the bulk heating effect will dominate and the edge effects can be ignored.  In this approximation, the heating is isotropic and 2/3 of the $E_{\rm DIH}$ will contribute to the MTE of the beam.  Assuming that the initial and DIH contributions can be added in quadrature, the core emittance becomes approximately:

\begin{equation}
    \epsilon_c \approx \frac{\sqrt{2\pi}}{4}\sigma_x \sqrt{\frac{\text{MTE}+\frac{2}{3}  E_{\rm DIH}}{mc^2}}
\end{equation}

In the case of the dc (NCRF) beamline with 0 meV MTE, the initial transverse size of the beam is 8.2 $\mu$m (2.6 $\mu$m) and the energy from DIH is 0.65 meV (1.4 meV), so the resulting core emittance due to $E_{\rm DIH}$ is 0.18 nm (0.070 nm).

To compare these results to those found in the previous sections, we must take into account the increase of the core emittance in the mean-field space charge simulation from 0, which is an effect of finite sampling. To do this, we will assume the effects of point-to-point space charge can be added in quadrature to the mean-field core emittance, analogously to equation (9) and as is valid for independent rms emittance contributions.  With this assumption, the core emittance contribution of point-to-point space charge, $\epsilon_{c,P2P}$, can be found through a quadrature subtraction of the mean-field core emittance from the PMP core emittance.  For the dc(NCRF) gun beamline, $\epsilon_{c,P2P}$ at the sample is .27 nm (.12 nm).  
67\% of $\epsilon_{c,P2P}$ at the sample in the dc case is explained by $E_{\rm DIH}$ (0.27 nm vs 0.18 nm), and in the rf case, 58\% of $\epsilon_{c,P2P}$ at the sample is determined by $E_{\rm DIH}$ (0.12 nm vs 0.070 nm).

For the rms transverse emittance contribution, we will use the intrinsic emittance of the beamline \cite{IvanMTE}:

\begin{equation}
    \epsilon_i = \sigma_x  \sqrt{\frac{\text{MTE}}{mc^2}}
\end{equation}

As with the core emittance, adding in 2/3 of the disorder induced heating energy, a modification is made to the intrinsic emittance equation:

\begin{equation}
    \epsilon_i \approx \sigma_x  \sqrt{\frac{\text{MTE}+\frac{2}{3}  E_{\rm DIH}}{mc^2}}
\end{equation}

In the dc (NCRF) beamline, the intrinsic 90\% transverse emittance including $E_{\rm DIH}$ is 0.27 nm (0.11 nm). We will compare this to the quadrature subtraction of the 90\% transverse emittance emittance of the mean-field simulation from the PMP simulation, which we will call $\epsilon_{P2P}$.  At the emittance minimum, $\epsilon_{P2P}$ in the dc (NCRF) beamline is 0.40 nm (0.18 nm).  The intrinsic emittance contribution from disorder induced heating in the dc (NCRF) beamline thus accounts for 68\% (61\%) of $\epsilon_{P2P}$.  The remaining difference can be attributed to the effect of large angle scatters which kick particles far from the beam center, seen as the tails of the emittance fraction curve in Fig. \ref{fig:CdcPF}.

\section{Conclusion}

In this work, we have shown that as photoemitted electron beam temperatures are made ever smaller, the effects of the point like nature of the Coulomb interaction become crucial to understanding photoinjector beam dynamics. We have introduced and benchmarked a simple method to compute the image force in a point-to-point beam dynamics simulations free of divergences and additional tuning parameters. Using this method, we have quantified Coulomb scattering effects on the beam phase space density in two UED beamline archetypes. For a photcathode with zero intrinsic emittance, the emittance of the beam in both an rf and dc UED beamline was larger by a factor of at least 2, and the core emittance is larger by a factor of at least 3 when compared to simulation on the same beamline but assuming mean-field space charge.  In addition, the energy released by disorder induced heating was calculated using the radial distribution function, and the heating was found to scale with the density to the power of 0.39 $\pm$ 0.03, close to the a simple theoretical estimate of 1/3, and was shown to be the dominant effect in both core and 90\%  rms emittance growth.

\section{acknowledgment}
This work was supported by the Center for Bright Beams, NSF PHY-1549132.

\appendix
\section{Warm Beam Comparison}
\label{Appendix:Warm}
In this section of the appendix, the same dc and NCRF UED beamlines were simulated using a initial beam MTE of 150 meV.  By doing so, we aim to show that the PMP and mean-field methods converge to the same result for a warm photocathode, and that Debye screen effectively mitigates point-to-point effects. 

\begin{figure}

\begin{subfigure}[htbp]{0.45\textwidth}
\centering
\includegraphics[width=\linewidth]{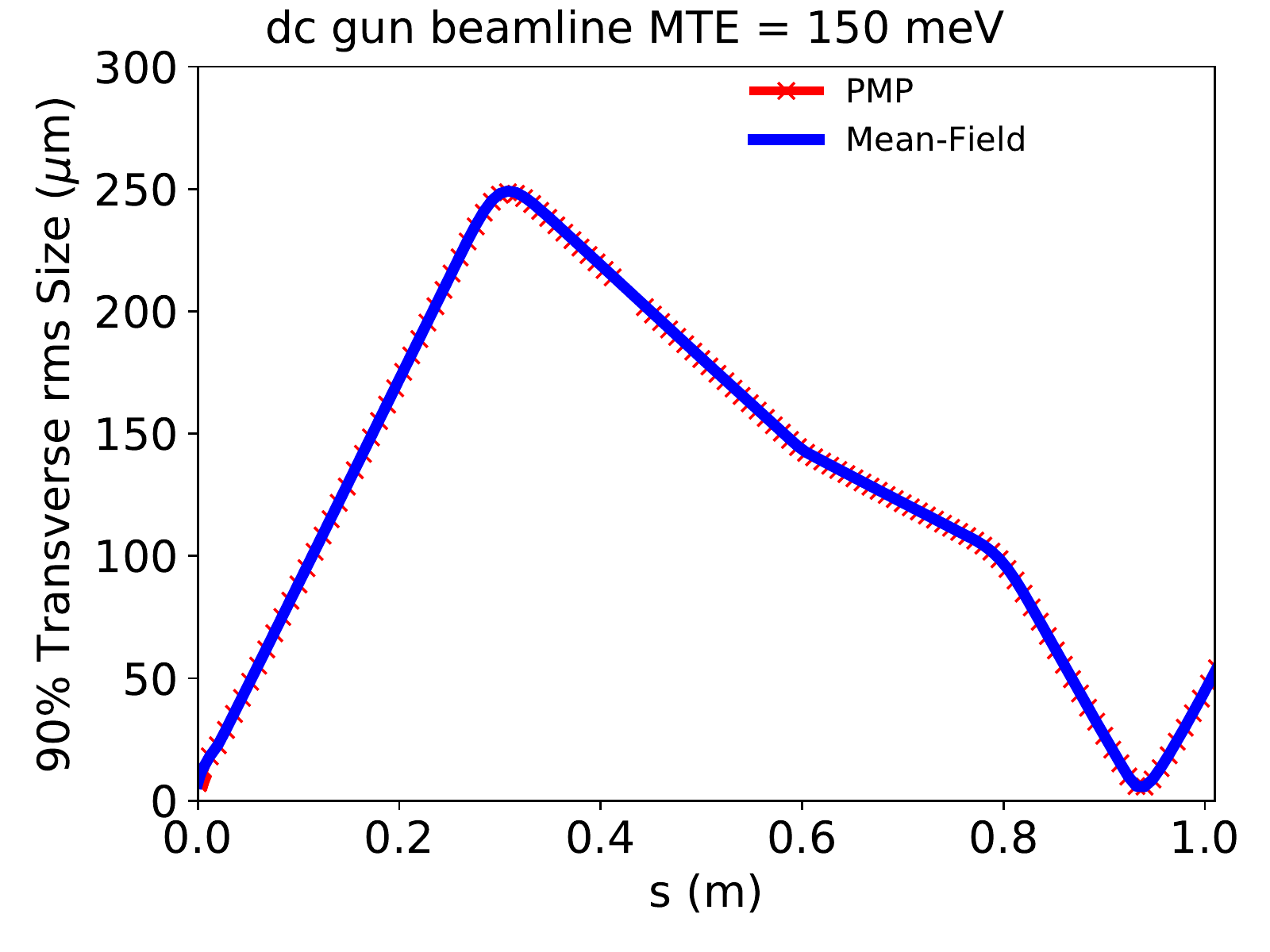}
\caption{}
\label{fig:150CdcSS}
\end{subfigure}

\begin{subfigure}[htp]{0.45\textwidth}
\centering
\includegraphics[width=\linewidth]{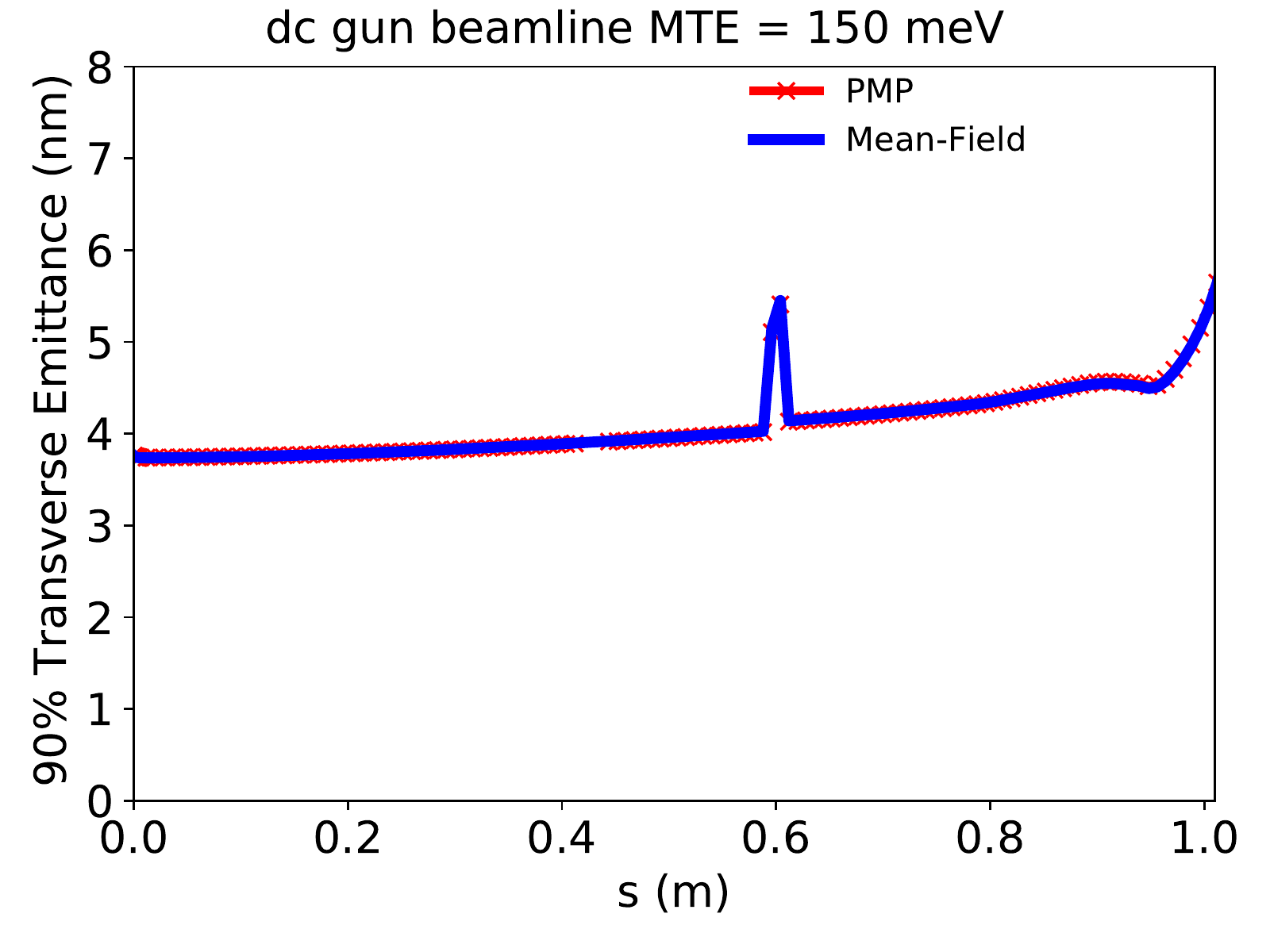}
\caption{}
\label{fig:150CdcE}
\end{subfigure}

\begin{subfigure}[htp]{0.45\textwidth}
\centering
\includegraphics[width=\linewidth]{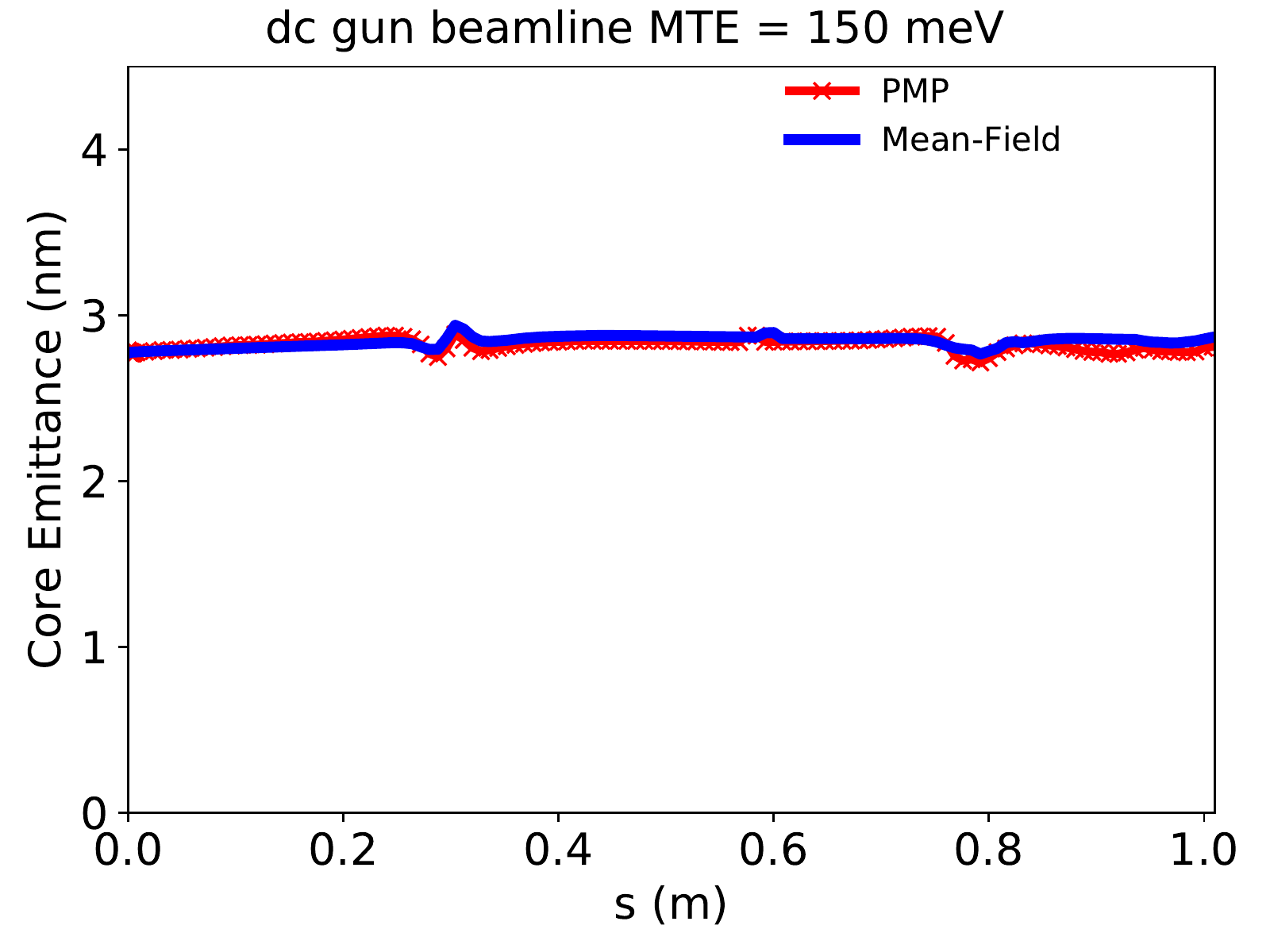}
\caption{}
\label{fig:150CdcCE}
\end{subfigure}

\caption{Spot size, transverse normalized rms emittance, and core emittance comparison between the PMP method and mean-field space charge simulations of the dc UED beamline with 150 meV MTE.}  
\end{figure}

\begin{figure}

\begin{subfigure}[htbp]{0.45\textwidth}
\centering
\includegraphics[width=\linewidth]{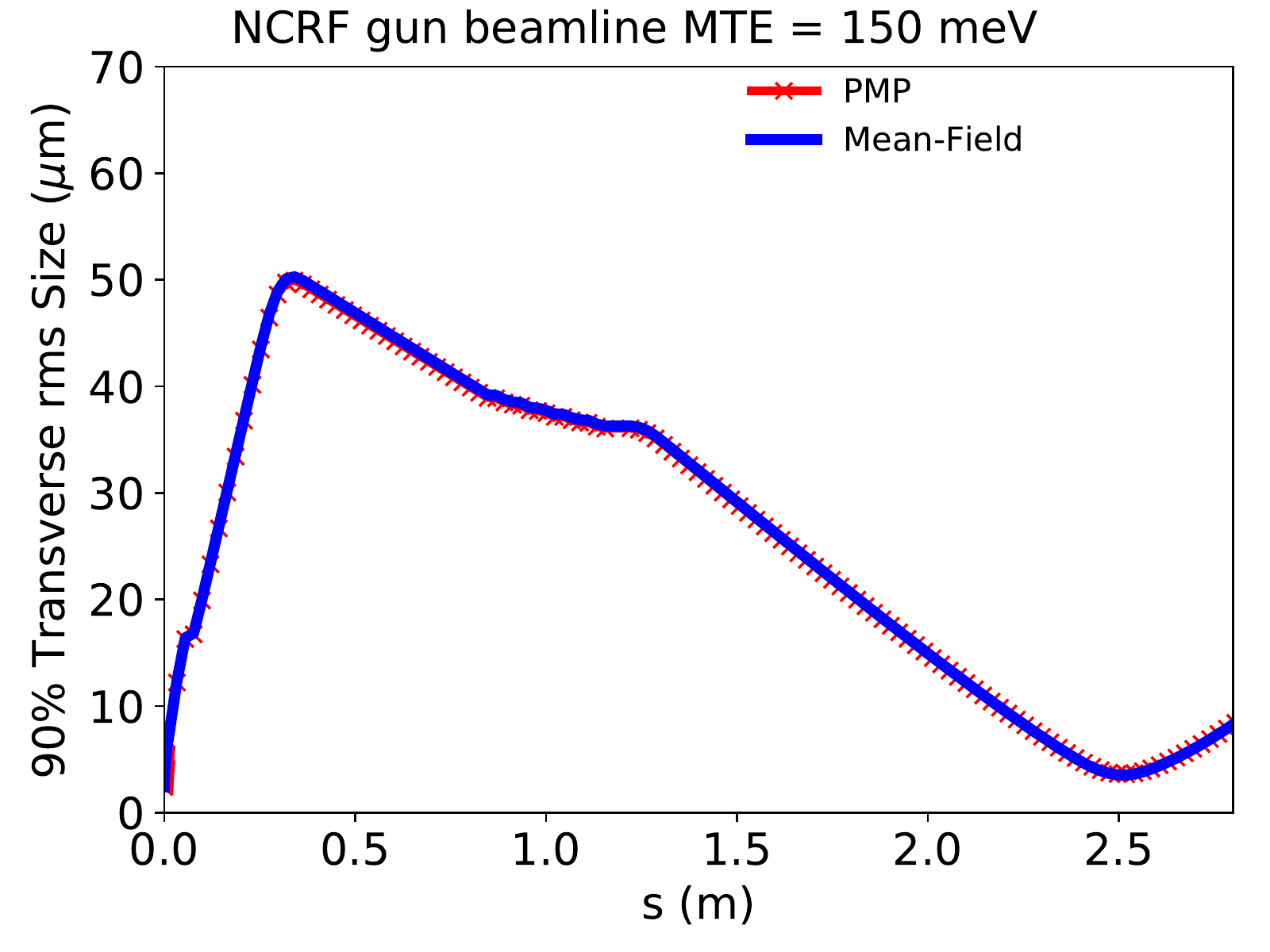}
\caption{}
\label{fig:150RFSS}
\end{subfigure}

\begin{subfigure}[htp]{0.45\textwidth}
\centering
\includegraphics[width=\linewidth]{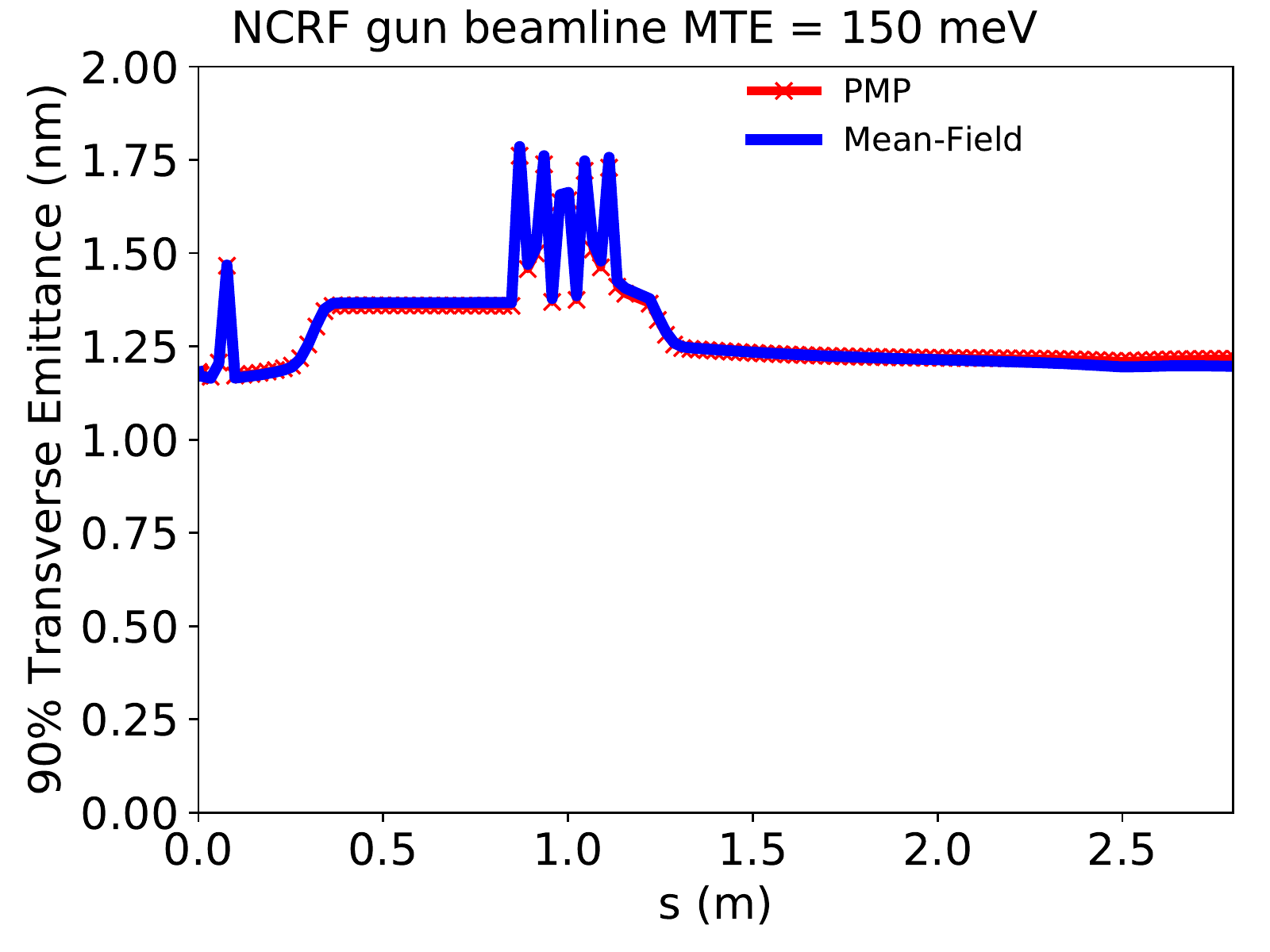}
\caption{}
\label{fig:150RFE}
\end{subfigure}

\begin{subfigure}[htp]{0.45\textwidth}
\centering
\includegraphics[width=\linewidth]{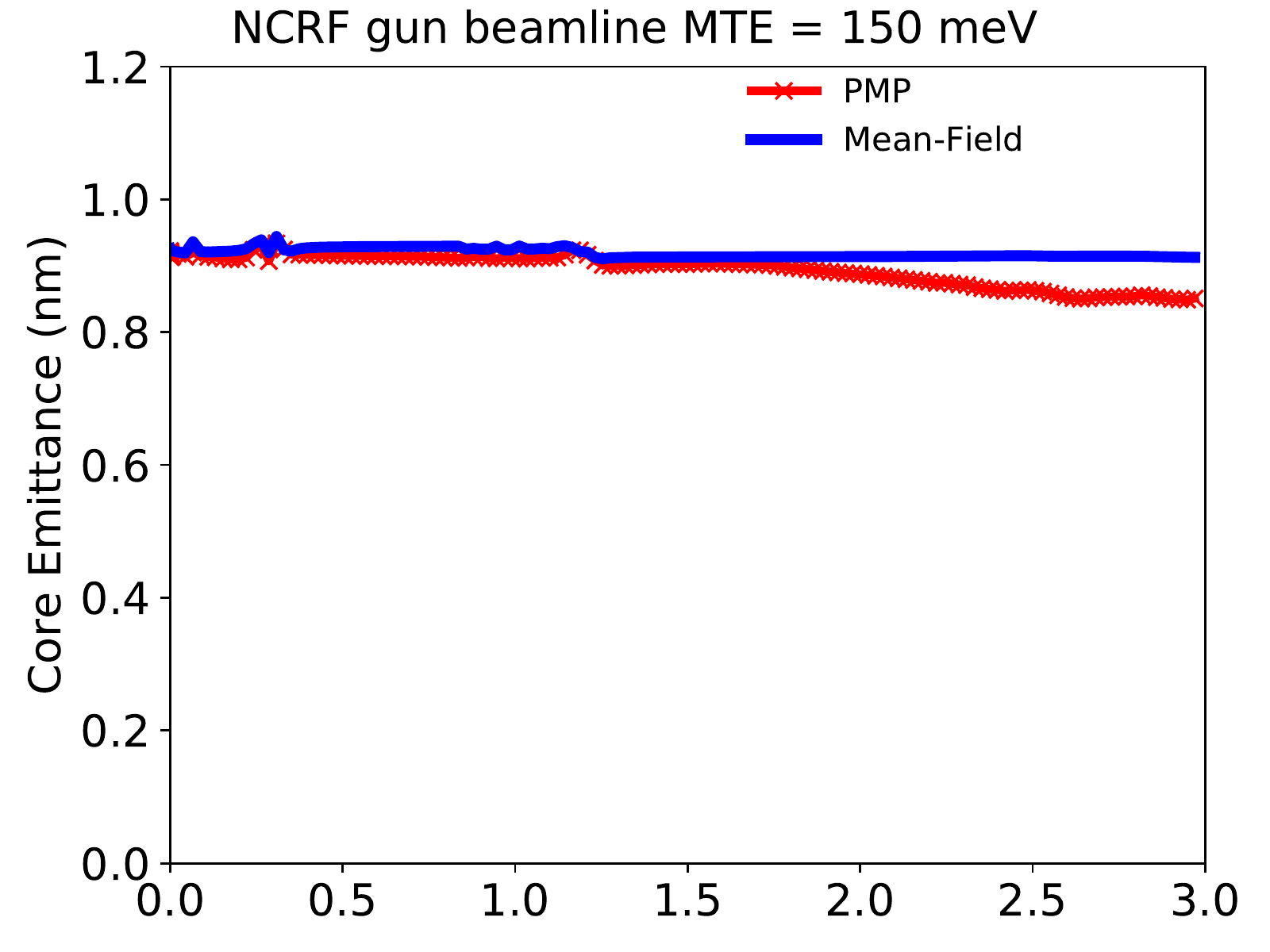}
\caption{}
\label{fig:RFCE150}
\end{subfigure}

\caption{Spot size, transverse normalized rms emittance, and core emittance comparison between the PMP method and mean-field space charge simulations of the NCRF UED beamline with 150 meV MTE.}  
\end{figure}

In Figs. \ref{fig:150CdcSS} and \ref{fig:150RFSS}, the spot size evolution is shown for the dc and NCRF UED beamlines respectively each with 150 meV MTE.  At this higher MTE, the difference between the point-to-point and mean-field spot size has been significantly reduced.  At the beams maximal size in the first solenoid, the deviation has been reduced from around 2 $\mu$m to .7 $\mu$m in the dc beamline and 1.2 $\mu$m to .5 $\mu$m in the NCRF beamline.  

The evolution of the transverse rms emittance for the dc and NCRF beamlines is shown in Figs. \ref{fig:150CdcE} and \ref{fig:150RFE} respectively.  With an MTE of 150 meV, there is no significant deviation in the transverse rms emittance between the implementations of space charge as observed at 0 meV.  This validates that phtocathodes with high emission temperatures can be successfully modelled without consideration of  point-to-point space charge effects.

The evolution of the core emittance at an MTE of 150 meV is shown in Figs. \ref{fig:150CdcCE} and \ref{fig:RFCE150} for the dc and NCRF UED beamlines respectively.  Outside of fluctuations near the solenoids the core emittance in all simulations are approximately constant.  No significant difference exists between the core emittance between point-to-point and mean-field simulations at 150 meV.

\section{Modified Image Charge Method}

The majority of this manuscript employs a mean-field model of the image force. In this appendix, we show this to be a valid approximation. To do this, we analytically investigate a  point-like image model and compare it to PMP.

We first aim to show that for particles much closer to the cathode compared to the average interparticle distance, the Coulombic repulsion force will be predominantly longitudinal, as the transverse fields from other charges and their images will largely cancel.

The transverse electric field from an electron a distance $d$ away from an infinite conducting plane and its image charge in cylindrical coordinates is given by:

\begin{equation}
E_r = \frac{1}{4\pi \epsilon_0}\left(\frac{-e r}{(r^2+(z-d)^2)^{3/2}}+\frac{e r}{(r^2+(z+d)^2)^{3/2}}\right)
\label{eq:EIC}
\end{equation}

 We first will consider the effects of particles far away from the cathode on particles which are recently emitted. At a position much closer to the cathode (z$\ll$d), this expression is approximately:

\begin{equation}
\label{eq:FFC}
E_r \approx \frac{1}{4\pi \epsilon_0}\frac{-3ezrd}{2(r^2+d^2)^{5/2}}
\end{equation}

We compare this to the longitudinal field in cylindrical coordinates:

\begin{equation}
E_z = \frac{1}{4\pi \epsilon_0}\left(\frac{-e (z-d)}{(r^2+(z-d)^2)^{3/2}}+\frac{e (z+d)}{(r^2+(z+d)^2)^{3/2}}\right)
\label{eq:EICL}
\end{equation}

In the same limit, $z \ll d$:
\begin{equation}
E_z \approx \frac{1}{4\pi \epsilon_0}\frac{2e d}{(r^2+d^2)^{3/2}}
\label{eq:EICL2}
\end{equation}

The magnitude of the ratio of the transverse and longitudinal electric fields in this limit of $z \ll d$ is:
\begin{equation}
 \left\lvert{\frac{E_r}{E_z}}\right\rvert  \approx \frac{3zr}{4(r^2+d^2)}
\end{equation}

This ratio tends to 0 for small or large $r$ and has a maximum at $r=d$ of:
\begin{equation}
 \left\lvert{\frac{E_r}{E_z}}\right\rvert_{max}  \approx \frac{3z}{8d}
\end{equation}
Because $z \ll d$ the transverse electric field will be much smaller than the longitudinal field, and thus proving the Coulombic repulsion force will be predominantly longitudinal.

For an electron beam with an average interparticle distance $a$, this will apply as long as $a \gg$ d.  Therefore, for particles that have just been emitted, if we are to model the effects of the cathode and to avoid divergent fields for small $z$, we can ignore transverse effects, and only need to find the time it takes for a particle to travel a longitudinal distance on the order of the interparticle distance, and its energy at that point.

From dynamical image charge theory, a semi-classical image potential can be computed \cite{DImC}, in the approximation that the electron has no velocity parallel to the conducting surface, the potential energy V is:
\begin{equation}
\begin{split}
    V(v,z)=-\frac{1}{4 \pi \epsilon_0}\frac{e^2\omega_p^2}{4v\omega_s}f\left(\frac{2z\omega_s}{v}\right),
   \\
    f(x)=\int^\infty_0\frac{e^{-\alpha x}}{1+\alpha^2}d\alpha
\end{split}
\end{equation}

where $\omega_p$ is the plasma frequency of the material, $\omega_p^2=2\omega_s^2$, $v$ is the velocity of the electron, and $z$ is its distance from the cathode.

The energy of an electron moving at a velocity $v$ a distance $z$ from the cathode with an applied electric field $E_z$ is thus:

\begin{equation}
    E = \frac{mv^2}{2}-\frac{1}{4 \pi \epsilon_0}\frac{e^2\omega_p^2}{4v\omega_s}f\left(\frac{2z\omega_s}{v}\right)-e E_z z
\end{equation}

The applied electric field will consist of the field from the gun as well as an approximation to the longitudinal effects of other particles.  The field from the particles in front of a given particle will be approximated as a uniformly charged cylinder with a transverse size, $R$, equal to that of the beam, and longitudinal size, $L$, equal to the initial bunch length multiplied by the beam fraction that has left the cathode previously.  The image charge contribution from other particles will be approximated as a positively charged cylinder with equal dimensions placed directly behind the particle.  With this, the equation for the applied electric field on the $j^{\rm th}$ particle emitted is:

\begin{equation}
    E_{z}(j) = E_{gun}(z) + \frac{e (j-1)}{\pi R^2 \epsilon_0}\left(1+\frac{R}{L(j)}-\sqrt{1+\frac{R^2}{L(j)^2}}\right)
\end{equation}
where $L(j)$ is the length of the bunch in front of the $j^{\rm th}$ particle.

These equations can be self-consistently solved to find the velocity and potential at any longitudinal position for a given energy $E$. Calculating the velocity at several locations near the cathode, the time it takes to reach a given distance can be calculated as well.  Because the transverse effects of space charge can be ignored in this regime, the external transverse field near the cathode can be used to approximate the transverse position of particle at a later distance, although we find this effect to be insignificant for the cases considered in this manuscript.

With this information, the beam can be initialized in simulation at a position away from the cathode divergence and thereafter can be modeled using a standard, point-like image charge method.  As long as the position is chosen to be sufficiently far from the cathode and less than the average particle distance, the resulting simulation should not depend heavily on the starting distance choice.

\begin{figure}[htp]
\centering
\includegraphics[width=220pt]{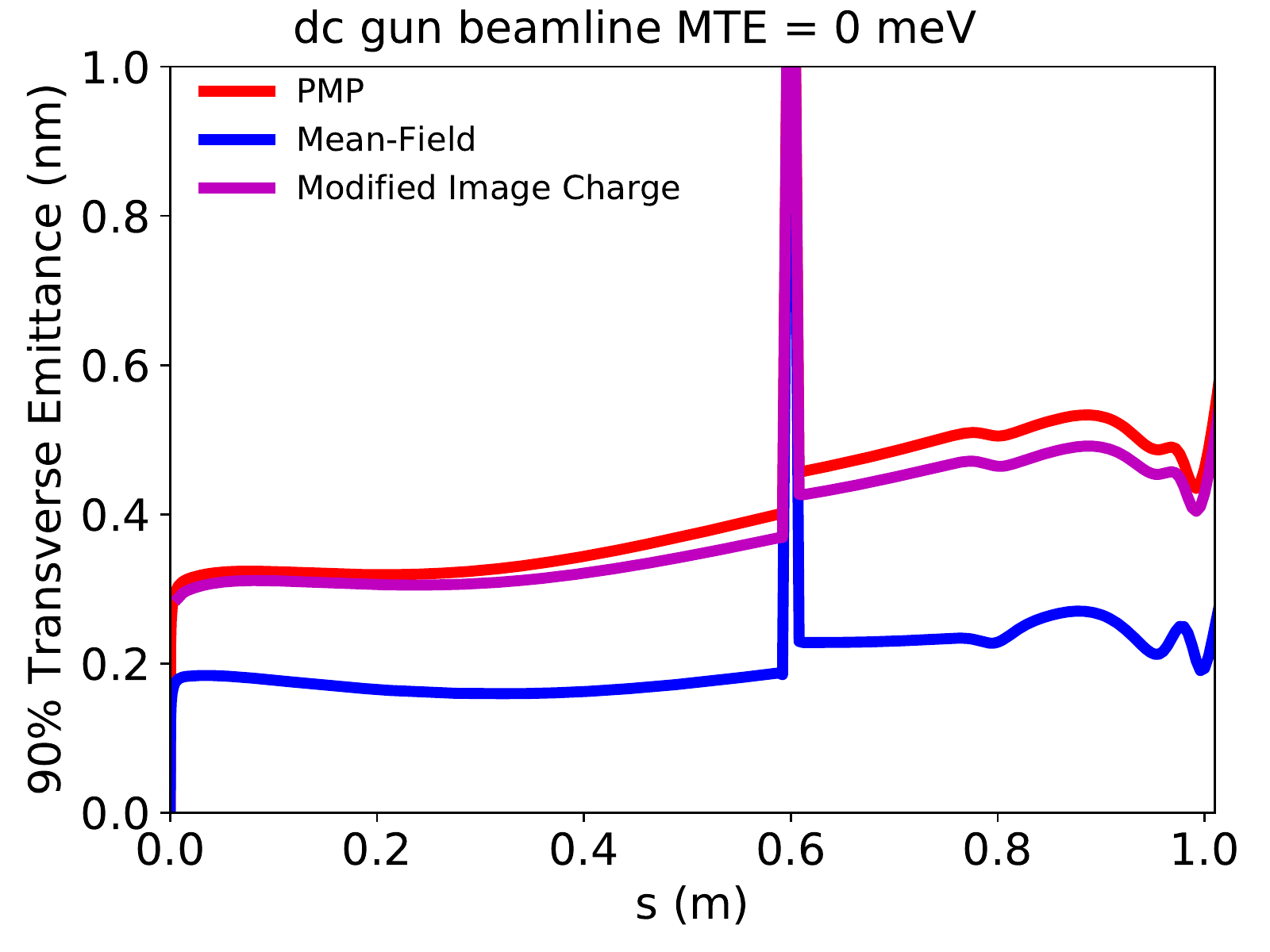}
\caption{Transverse normalized rms emittance comparison for point-to-point, mean-field, and modified image charge simulations of the dc UED beamline with 0 meV MTE.}  
\label{fig:ICdcE}
\end{figure}

\begin{figure}[htbp]
\centering
\includegraphics[width=220pt]{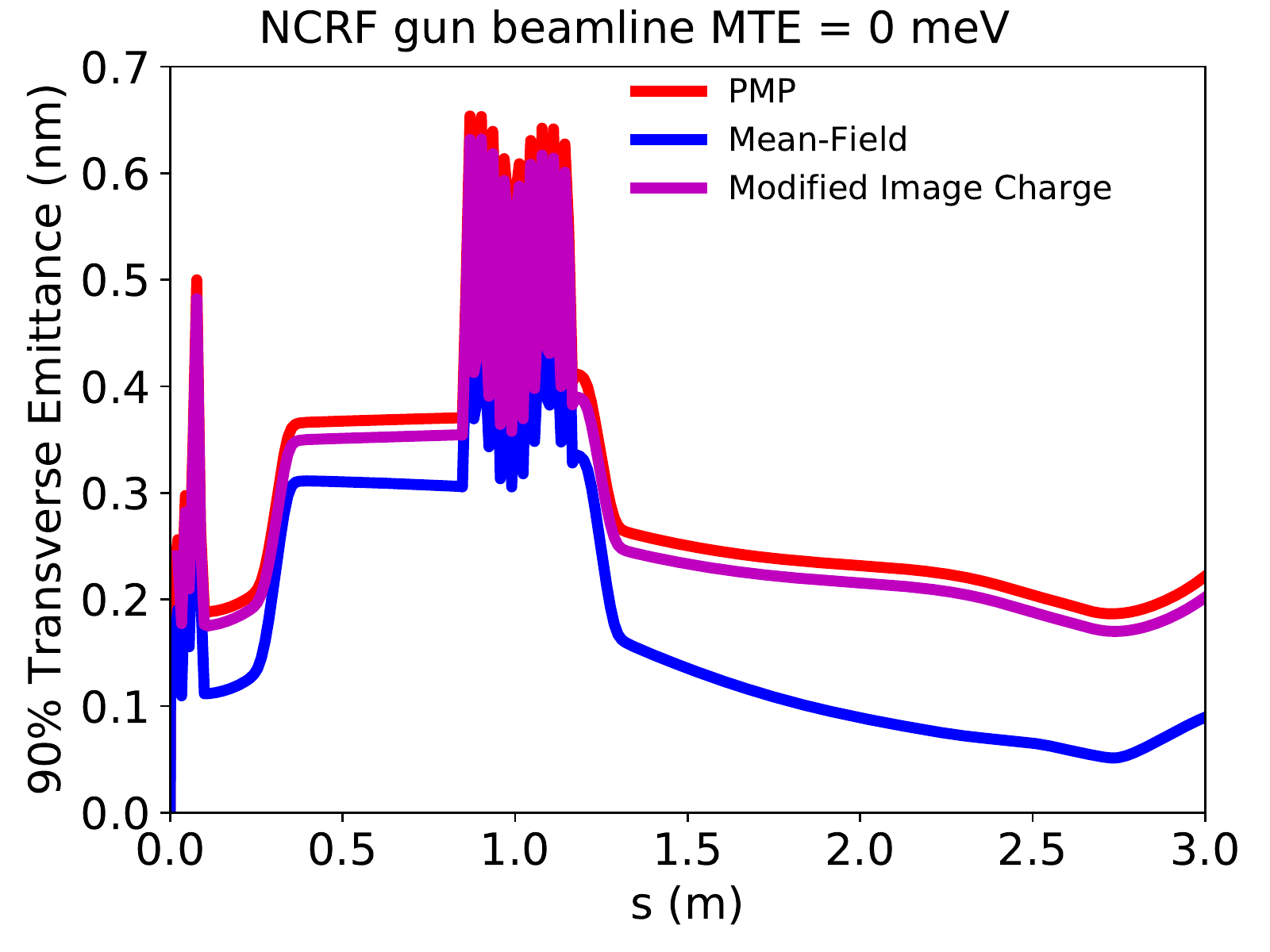}
\caption{Transverse normalized rms emittance comparison for point-to-point, mean-field, and modified image charge simulations for the NCRF UED beamline with 0 meV MTE.}  
\label{fig:IRFE}
\end{figure}

Through this analysis, we have shown that the image charges have little effect on the transverse dynamics of particles until a distance away from the cathode on the order of the average interparticle distance.  Because point-to-point effects are most important for interactions at a distance less than the average interparticle distance, a mean-field implementation of the image charge, such as is done through the PMP method used in this paper, should produce correct results as long as the energy modulation of the produced particles by their images is correct.

In Figs. \ref{fig:ICdcE} and \ref{fig:IRFE}, the 0 meV MTE transverse normalized rms emittance results for the DC and NCRF beamline are shown respectively, now including results from the modified image charge method discussed in this appendix.  $E$ was chosen such that the minimum kinetic energy of an electron was 5 meV, on the order of the smallest MTE measured today.  The particles were started at a distance 0.5 $\mu$m away from the cathode.  Deviations in the emittance from the PMP method are less than 10\% throughout the simulations.

\bibliography{P2PUED}
\end{document}